\documentclass{aa}

\usepackage{graphicx}
\usepackage{txfonts}
\usepackage{siunitx}
\usepackage[colorlinks]{hyperref}
\usepackage{tikz}

\let\oldsim\sim 
\renewcommand{\sim}{{\oldsim}}
\DeclareSIUnit \electron {\ensuremath{\mathrm{e}^-}}
\DeclareSIUnit \pix {pix}
\DeclareSIUnit \rms {\ensuremath{_\mathrm{RMS}}}
\DeclareSIUnit \mag {\ensuremath{\mathrm{mag}}}

\newcommand*\annotatedFigureBoxCustom[8]{\draw[#5,thick,rounded corners] (#1) rectangle (#2);\node at (#4) [fill=#6,thick,shape=circle,draw=#7,inner sep=2pt,font=\sffamily,text=#8] {\textbf{#3}};}
\newcommand*\annotatedFigureBox[4]{\annotatedFigureBoxCustom{#1}{#2}{#3}{#4}{white}{white}{black}{black}}

\newenvironment {annotatedFigure}[1]{\centering\begin{tikzpicture}
\node[anchor=south west,inner sep=0] (image) at (0,0) { #1};\begin{scope}[x={(image.south east)},y={(image.north west)}]}{\end{scope}\end{tikzpicture}}
\usepackage{amstext}

\begin{document}

\title{NAOMI: the adaptive optics system\\of the Auxiliary Telescopes of the VLTI}

\subtitle{}

\author{  J.~Woillez\inst{1}\thanks{E-mail: jwoillez@eso.org} \and
        J.A.~Abad\inst{1} \and
          R.~Abuter\inst{1} \and
          E.~Aller~Carpentier\inst{1} \and
          J.~Alonso\inst{2} \and
          L.~Andolfato\inst{1} \and
          P.~Barriga\inst{1} \and
       J.-P.~Berger\inst{3} \and
       J.-L.~Beuzit\inst{3} \and
          H.~Bonnet\inst{1} \and
          G.~Bourdarot\inst{3} \and
          P.~Bourget\inst{2} \and
          R.~Brast\inst{1} \and
          L.~Caniguante\inst{2} \and
          E.~Cottalorda\inst{3} \and
          P.~Darré\inst{1} \and
          B.~Delabre\inst{1} \and
          A.~Delboulbé\inst{3} \and
          F.~Delplancke-Ströbele\inst{1} \and
          R.~Dembet\inst{1} \and
          R.~Donaldson\inst{1} \and
          R.~Dorn\inst{1} \and
          J.~Dupeyron\inst{1} \and
          C.~Dupuy\inst{1} \and
          S.~Egner\inst{1} \and
          F.~Eisenhauer\inst{6} \and
          G.~Fischer\inst{1} \and
          C.~Frank\inst{1} \and
          E.~Fuenteseca\inst{2} \and
          P.~Gitton\inst{2} \and
          F.~Gonté\inst{1} \and
          T.~Guerlet\inst{1} \and
          S.~Guieu\inst{3} \and
          P.~Gutierrez\inst{1} \and
          P.~Haguenauer\inst{1} \and
          A.~Haimerl\inst{1} \and
          X.~Haubois\inst{2} \and
          C.~Heritier\inst{1} \and
          S.~Huber\inst{1} \and
          N.~Hubin\inst{1} \and
          P.~Jolley\inst{1} \and
          L.~Jocou\inst{3} \and
       J.-P.~Kirchbauer\inst{1} \and
          J.~Kolb\inst{2} \and
          J.~Kosmalski\inst{1} \and
          P.~Krempl\inst{4} \and
       J.-B.~Le~Bouquin\inst{3} \and
          M.~Le~Louarn\inst{1} \and
          P.~Lilley\inst{1} \and
          B.~Lopez\inst{7} \and
          Y.~Magnard\inst{3} \and
          S.~Mclay\inst{1} \and
          A.~Meilland\inst{7} \and
          A.~Meister\inst{2} \and
          A.~Merand\inst{1} \and
          T.~Moulin\inst{3} \and
          L.~Pasquini\inst{1} \and
          J.~Paufique\inst{1} \and
          I.~Percheron\inst{1} \and
          L.~Pettazzi\inst{1} \and
          O.~Pfuhl\inst{6} \and
          D.~Phan\inst{1} \and
          W.~Pirani\inst{1} \and
          J.~Quentin\inst{1} \and
          A.~Rakich\inst{1} \and
          R.~Ridings\inst{1} \and
          M.~Riedel\inst{1} \and
          J.~Reyes\inst{1} \and
          S.~Rochat\inst{3} \and
          G.~Santos~Tomás\inst{1} \and
          C.~Schmid\inst{1} \and
          N.~Schuhler\inst{2} \and
          P.~Shchekaturov\inst{5} \and
          M.~Seidel\inst{1} \and
          C.~Soenke\inst{1} \and
          E.~Stadler\inst{3} \and
          C.~Stephan\inst{2} \and
          M.~Suárez\inst{1} \and
          M.~Todorovic\inst{1} \and
          G.~Valdes\inst{2} \and
          C.~Verinaud\inst{1} \and
          G.~Zins\inst{3} \and
          S.~Zúñiga-Fernández\inst{2,8,9}
        }
\institute{European Southern Observatory Headquarters,
           Karl-Schwarzschild-Straße 2, 85748 Garching bei München, Germany
           \and
           European Southern Observatory Vitacura,
           Alonso de Córdova 3107, Vitacura, Casilla 19001, Santiago de Chile, Chile
           \and
           Université Grenoble Alpes,
           CNRS, IPAG, F-38000 Grenoble, France
           \and
           KRP Mechatec GmbH,
           Boltzmannstraße 2, 85748 Garching bei München, Germany
           \and 
           Pactum LTD,
           PO Box 71035, London W4 9GZ, United Kingdom
           \and
           Max Planck Institute for extraterrestrial Physics, 
           Giessenbachstr., 85741 Garching, Germany
           \and
           Lab. Lagrange, Univ. Côte d'Azur, Observatoire de la Côte d'Azur, CNRS, France
           \and
           Universidad de Valparaíso, Instituto de Física y Astronomía (IFA), Avenida Gran Bretaña 1111, Casilla 5030, Valparaíso, Chile
           \and 
           Núcleo Milenio de Formación Planetaria (NPF), Valparaíso, Chile
          }

\idline{A\&A forthcoming}
\doi{https://doi.org/10.1051/0004-6361/201935890}
\date{Received: 15 May 2019 / Accepted: 02 July 2019}

\abstract
   {The tip-tilt stabilisation system of the \SI{1.8}{\meter} Auxiliary Telescopes of the Very Large Telescope Interferometer was never dimensioned for robust fringe tracking, except when atmospheric seeing conditions are excellent.}
   {Increasing the level of wavefront correction at the telescopes is expected to improve the coupling into the single-mode fibres of the instruments, and enable robust fringe tracking even in degraded conditions.}
   {We deployed a new adaptive optics module for interferometry (NAOMI) on the Auxiliary Telescopes.}
   {We present its design, performance, and effect on the observations that are carried out with the interferometric instruments.}
   {}


\maketitle

\section{Introduction}

The \SI{1.8}{\meter} Auxiliary Telescopes \citep[ATs,][]{Koehler+2002} of the Very Large Telescope Interferometer (VLTI) achieved first fringes in 2005 \citep{Koehler+2005} and started operation in $2006$.
Since the beginning, the telescopes have been equipped with a fast tip-tilt correction system: a STRAP tip-tilt sensor \citep{Bonaccini+1997} at the coudé focus paired with a fast tip-tilt mirror at the M6 location in the coudé train, even though the original intent was to equip them with a higher order adaptive optics (AO) system \citep{Beckers1990}.

For a median seeing of \SI{1.1}{\arcsec} at the level of the ATs\footnote{The median free-atmosphere seeing of Paranal is \SI{0.8}{\arcsec} \citep{Sarazin+2008}. At the level of the ATs, an additional ground-layer seeing must be taken into account, resulting in an \SI{1.1}{\arcsec} median seeing.}, a tip-tilt correction on an \SI{1.8}{\meter} telescope is somewhat cost-optimal, in terms of the $V^2$ signal-to-noise ratio (S/N), at L band, where $D{\approx}3r_0$ \citep{Keen+2001,Haniff2007}.
This $D/r_0\le3$ criteria for a tip-tilt-corrected telescope is however not adequate to co-phase the array.
For a fringe tracker to work properly, it must be capable of continuously measuring the fringe phase with a $\mathrm{S/N}_\phi > 1$.
Even at $D{\approx}r_0$, a single-mode fringe tracker suffers from significant flux dropouts.
Consequently, significantly brighter objects must be observed, which affects the co-phasing limiting magnitude of the interferometer.
Theoretical studies of the single-mode flux in partial AO correction \citep{Tatulli+2010} confirm that this behaviour is observed on the VLTI.
They explain in part why AMBER under-performed with its H-band FINITO fringe tracker \citep{Bonnet+2006, Merand+2012}.

The tip-tilt correction system also made all near-infrared VLTI observations with the ATs very sensitive to degraded seeing conditions.
Even more frustrating, excellent seeing episodes on Paranal are usually accompanied by low wind speed conditions.
Then, slow and strong turbulence develops inside the domes of the ATs, to a point where the delivered beams become multi-speckle in the near-infrared.
The excellent seeing nights therefore become unusable \citep{Woillez+2018} by a single-mode instrument like PIONIER \citep{LeBouquin+2011}, and even more by GRAVITY \citep{GravityCollaboration+2017} and its fringe tracker \citep{Lacour+2019}.

To remedy this situation, the AO project NAOMI for the ATs was officially launched in 2008 inside ESO.
In competition with the second-generation VLTI instruments GRAVITY, MATISSE \citep{Lopez+2014}, and especially PRIMA \citep{Delplancke2008}, the project only acquired momentum in 2016. After the upgrade of the interferometer for the second-generation instruments was completed \citep{Woillez+2015}, the GRAVITY instrument was installed, and IPAG became a partner in the project.
A final design review was passed in early 2017.
Construction and extensive laboratory tests continued at ESO Headquarters, until shipment to Chile in mid-2018.
The installation on all four ATs started at the beginning of September 2018, and the upgraded AT array was successfully returned to operations $\sim2.5$ months later.

In this paper, we give an overview of the NAOMI design (Sect.~\ref{Sec:DesignOverview}). We present the on-sky performance of the AO system (Sect.~\ref{Sec:Performance}) and analyse its effect on the VLTI and its instruments (Sect.~\ref{Sec:ImpactOnVlti}).

\section{Design overview}\label{Sec:DesignOverview}

In this section, we present an overview of the NAOMI design.
The first three parts present the following three main components: the deformable mirror (\ref{SSec:DeformableMirror}), the wavefront sensor (\ref{SSec:WavefrontSensor}), and the real-time controller (\ref{SSec:RealTimeController}). The last three parts cover additional capability and performance aspects: the acquisition procedure (\ref{SSec:AcquisitionProcedure}), pupil registration (\ref{SSec:PupilRegistration}), and chopping (\ref{SSec:Chopping}).
A summary of the key parameters of the NAOMI system are given in Table~\ref{Tab:KeyParameters}.

\begin{table}
    \centering
    \begin{tabular}{rl}
        \hline
        \hline
        \multicolumn{2}{c}{\ref{SSec:DeformableMirror} Deformable Mirror (ALPAO DM241)} \\
        \hline
        Tip-tilt stroke                 & $>$ \SI{40}{\micro\meter} (PtV WF) \\
        3x3 actuators stroke            & \SI{25}{\micro\meter} (PtV WF) \\
        Pupil size                      & \SI{28}{\milli\meter} \\
        Actuator counts                 & 241 total, $\sim145$ effective \\
        Actuator per sub-aperture       & $\sim4$ \\
        Number of controlled modes      & 14 Zernikes (after piston)\\
        Settling time                   & <\SI{1.6}{\milli\second} \\
        \hline
        \hline
        \multicolumn{2}{c}{\ref{SSec:WavefrontSensor} Wavefront Sensor (ANDOR iXon Ultra 897)} \\
        \hline
        Type                            & Shack-Hartmann \\
        Sub-apertures                   & Square - 4x4 array \\
        Nb of valid sub-apertures       & 12 \\
        Pixel scale                     & \SI{0.375}{\arcsec\per\pix} \\
        Sub-aperture acquisition        & $16\times16$ or \SI{6}{\arcsec}$\times$\SI{6}{\arcsec} \\
        Sub-aperture closed loop        &  $6\times6$  or \SI{2.25}{\arcsec}$\times$\SI{2.25}{\arcsec} \\
        Lenslet array pupil diameter    & \SI{2.05}{\milli\meter} \\
        Lenslet array focal length      & \SI{20}{\milli\meter} \\
        Readout noise                   & \SI{0.32}{\electron} (sensitive mode) \\
        Clock-induced charges           & \SI[per-mode=symbol]{0.002}{\electron\per\pix} \\
        Pixel size                      & \SI{32}{\micro\meter} ($2\times2$ binned) \\
        Shortest integration time       & \SI{1.731}{\milli\second} \\
        Pixel count                     & $64\times64$ \\
        Wavelength                      & \SIrange{450}{900}{\nano\meter} \\
      
        \hline
        \hline
        \multicolumn{2}{c}{\ref{SSec:RealTimeController} Real-Time controller (SPARTA Light)} \\
        \hline
        Loop Frequency                  & \SIrange{500}{50}{\hertz} \\
        Controller latency              & <\SI{1}{\milli\second} \\
        Full loop delay                 & \SI{4.6}{\milli\second} \\
        Reconstructor update            & <\SI{0.2}{\hertz} \\
        Slope measurement               & Weighted Centre of Gravity \\ 
        Wavefront reconstruction        & Modal MVM \\
        Additional capabilities         & Chopping, vibration control \\
        \hline
        \hline
    \end{tabular}
    \caption{Summary of the key parameters of the NAOMI AO system.}
    \label{Tab:KeyParameters}
\end{table}

\begin{figure}
    \begin{annotatedFigure}
        {\includegraphics[width=\linewidth]{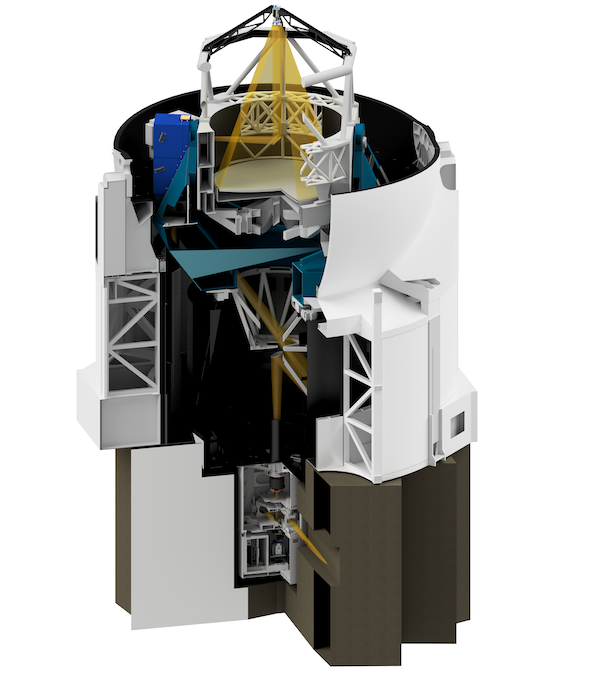}}
        \annotatedFigureBox{0.2135,0.666}{0.3749,0.856}{A}{0.2135,0.856}
        \annotatedFigureBox{0.3703,0.1089}{0.5427,0.346}{B}{0.3703,0.1089}
    \end{annotatedFigure}
    \caption{3D view of an AT; the NAOMI components are highlighted.
                \textbf{A}: Deformable mirror at the right Nasmyth.
                \textbf{B}: Wavefront sensor in the relay optics structure under the telescope.}
    \label{Fig:ViewAuxiliaryTelescope}
\end{figure}

\subsection{Deformable mirror}\label{SSec:DeformableMirror}

\begin{figure}
    \centering
    \begin{annotatedFigure}
        {\includegraphics[width=1.0\linewidth]{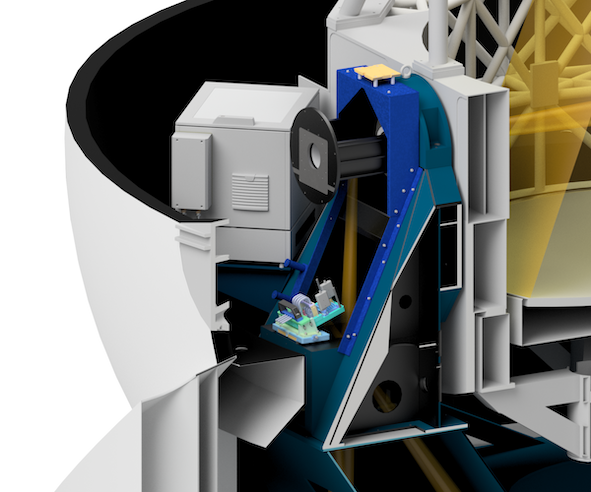}}
        \annotatedFigureBox{0.44,0.28}{0.60,0.45}{A}{0.44,0.28}
        \annotatedFigureBox{0.28,0.46}{0.55,0.84}{B}{0.28,0.84}
    \end{annotatedFigure}
    \caption{3D view of the right Nasmyth of the AT showing the deformable mirror and drive electronics at the M6 location. \textbf{A}: Deformable mirror mounted inside the large stroke quasi-static tip-tilt mount. \textbf{B}: M6 cabinet containing the deformable mirror and quasi-static mount drive electronics.}
    \label{Fig:ViewRightNasmyth}
\end{figure}

NAOMI uses an ALPAO DM241 deformable mirror (DM) with 241 actuators over a \SI{37.5}{\milli\meter} clear aperture.
It is installed in a pupil plane at the position of M6 in the coudé train.
As shown in Fig.~\ref{Fig:WfsDmGeometry}, the \SI{28}{\milli\meter} pupil size is smaller than the DM clear aperture, which makes the effective actuator count closer to $\sim145$.
This size mismatch is the result of our choice of an off-the-shelf DM and our decision to keep the coudé train of the telescope unchanged.
Its datasheet tip-tilt stroke of \SI{40}{\micro\meter} wavefront, equivalent to \SI{4.6}{\arcsecond} on sky, is well adapted to chopping directly with the DM, provided that the chopping stroke is optimised (see Sect. \ref{SSec:Chopping}).
To maximise the stroke further, the DM is mounted inside a large stroke gimbal tip-tilt mount, the so-called quasi-static mount (QSM).
Used for the alignment of the DM, the QSM absorbs known temperature-induced tip-tilt flexures of the coudé train.
The control electronics for the DM and the QSM are installed inside a cabinet that is mounted next to the M6 location (see Fig.~\ref{Fig:ViewRightNasmyth}).

Because the relative orientation of the DM and WFS changes with the azimuth and derotator angles, a rotation-stable Zernike basis is chosen as the control space, and derotation is performed in software by rotating the control matrix.
Based on the chosen wavefront sensor geometry, performance measurements show an optimum at 14 controlled Zernikes.
The mode-to-DM (M2DM) matrix that converts the \si{\micro\meter\rms}-normalised Zernikes into DM commands is derived from the measured DM influence functions.
Because of the high actuator count inside the pupil, the Zernike modes can be reproduced with high fidelity,
keeping aliasing effects to a minimum as the DM rotates with respect to the WFS.

A settling time of \SI{1.5}{\milli\second} makes this DM fast to a point where strong \SIrange{700}{900}{\hertz} resonances must be attenuated by smoothing the command sent by the RTC (see \ref{SSec:RealTimeController}).
While exploring further the dynamical properties of the DM, it was discovered that the NAOMI modes, even though specified piston-free over the NAOMI pupil, were exciting a global piston resonance of the membrane at \SI{500}{\hertz}.
This effect was efficiently mitigated by making the modes piston-free over both the NAOMI pupil and the full DM aperture.
Even though only $\sim145$ actuators contribute to the wavefront in the pupil, all 241 are used to control the piston.
An illustration of the resulting defocus mode is shown in Fig.~\ref{Fig:DoublePistonFreeModes}.

The number of controlled modes (14 Zernikes) is significantly lower than the actuator count, even when compared to the $\sim145$ inside the pupil.
This makes it impossible for NAOMI itself to directly measure and control the high-order flatness of the DM.
With limited knowledge of the long-term and temperature-dependent evolution of the high-order flat, tools were specifically developed to help with the maintenance of the system.
A re-calibration bench was developed to regularly re-measure the influence functions and flat at various temperatures; and to help with this process, the DM and QSM assemblies were designed to be easily removable from the AT.
In addition, a technique was developed to measure up to 120 Zernikes by shifting the WFS relative to the DM with the pupil control unit \citep{Oberti+2019}.
This allows monitoring, in day-time, on the telescope Nasmyth light source, most of the high-order flat ($\sim80\%$ of the intra-pupil actuator count) without moving the DM from the telescope to the re-calibration bench.

The version of ALPAO's DM241 used for NAOMI is subject to a creep effect: probably a slow relaxation of the material used for the actuator springs.
The creep appears as an additional command, increasing with time, and proportional to the initial command.
To keep the impact of creep to a minimum, the temporal average of the DM position should stay close to zero, which is the case when the chopping stroke is optimised, as explained in section \ref{SSec:Chopping}.

The response of the DM was known to depend on the temperature, therefore a temperature sensor was initially installed on the DM housing, with the intent to re-scale the M2DM matrix to compensate for the variable DM gain.
After it was deployed on the telescope in an environment where the ambient temperature fluctuates significantly, the DM gain appeared rather decorrelated from the housing temperature, as shown by unexpected transfer functions and interaction matrix measurements.
The temperature of the spring element inside the DM would probably have been a better proxy for the DM gain but remained inaccessible.
Therefore, the acquisition procedure (see Sec.~\ref{SSec:AcquisitionProcedure}) was modified to include an on-sky estimation of the DM gain.
A modulation is injected on the two coma modes (Z7 and Z8) while the AO loop is closed on all other modes.
The coma response on the wavefront sensor yields the DM gain.

With a properly scaled M2DM, the DM behaves like a piston-free modal actuator (see \ref{SSec:RealTimeController}).
More details on the properties and characterisation of these DMs can be found in \citet{LeBouquin+2018} and \citet{Haguenauer+2019}.

\begin{figure}
    \centering
    \includegraphics[width=0.6\linewidth]{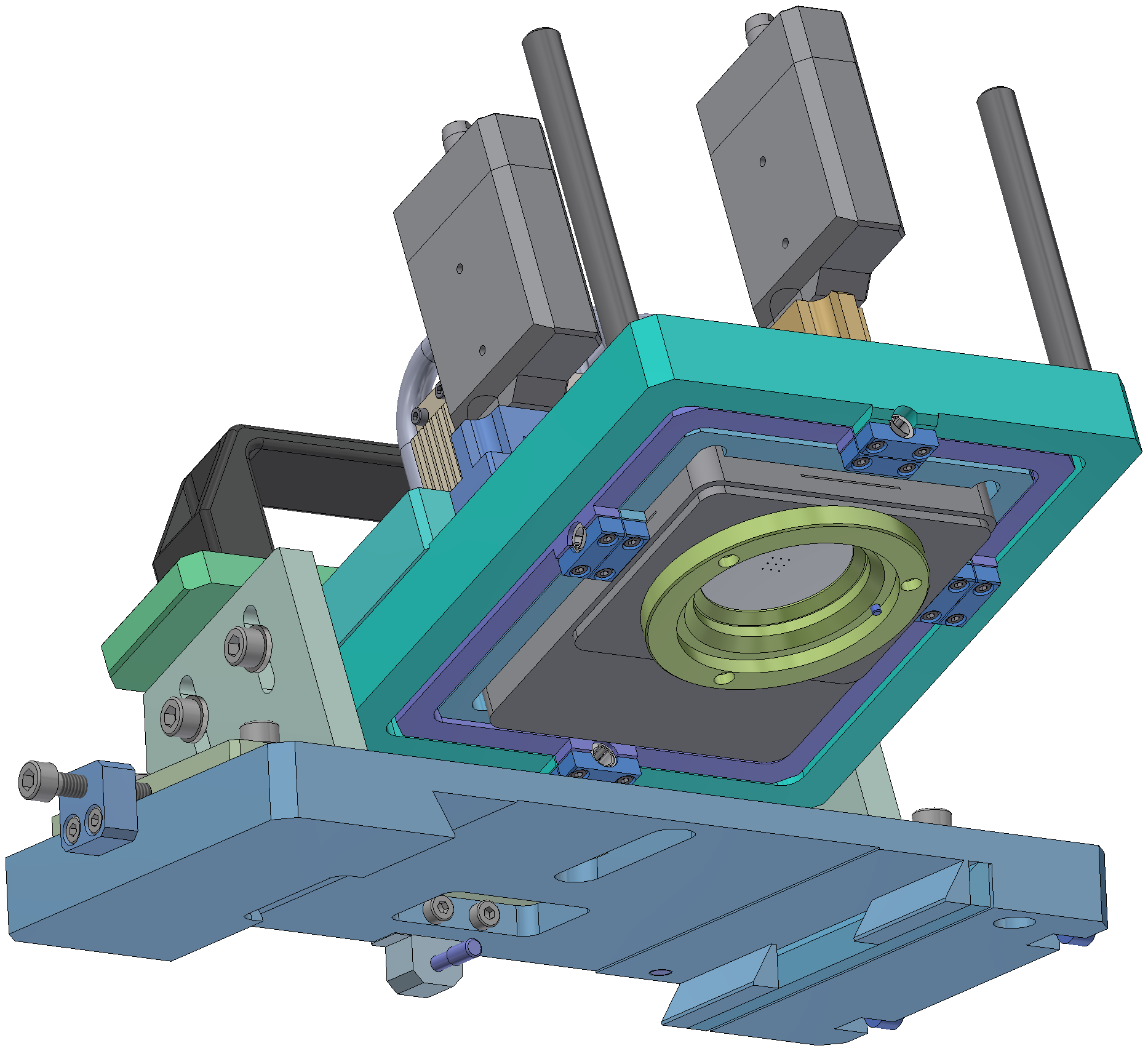}
    \caption{3D view of the corrective optics. The ALPAO DM241 is mounted in a large stroke gimbal tip-tilt mount. The assembly, installed inside the right Nasmyth assembly, can easily be removed when/if a re-calibration of the DM is needed.}
    \label{Fig:ViewCorrectiveOptics}
\end{figure}

\begin{figure}
    \centering
    \includegraphics[width=\linewidth]{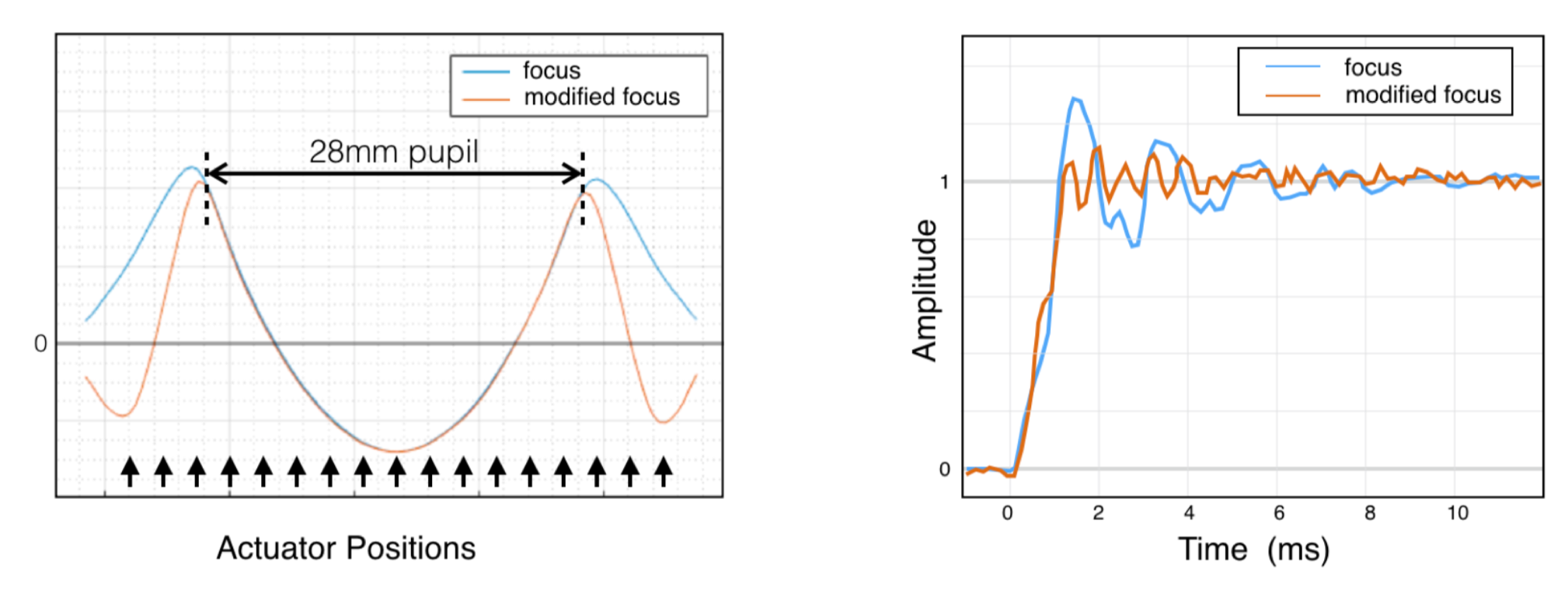}
    \caption{Radial modes are designed piston-free inside the \SI{28}{\milli\meter} pupil, and over the full clear aperture. Left: Profile of the defocus mode. Right: When the defocus mode is also piston-free over the full clear aperture, the \SI{500}{\hertz} resonance is not excited as much.}
    \label{Fig:DoublePistonFreeModes}
\end{figure}

\begin{figure}
    \centering
    \includegraphics[width=\linewidth]{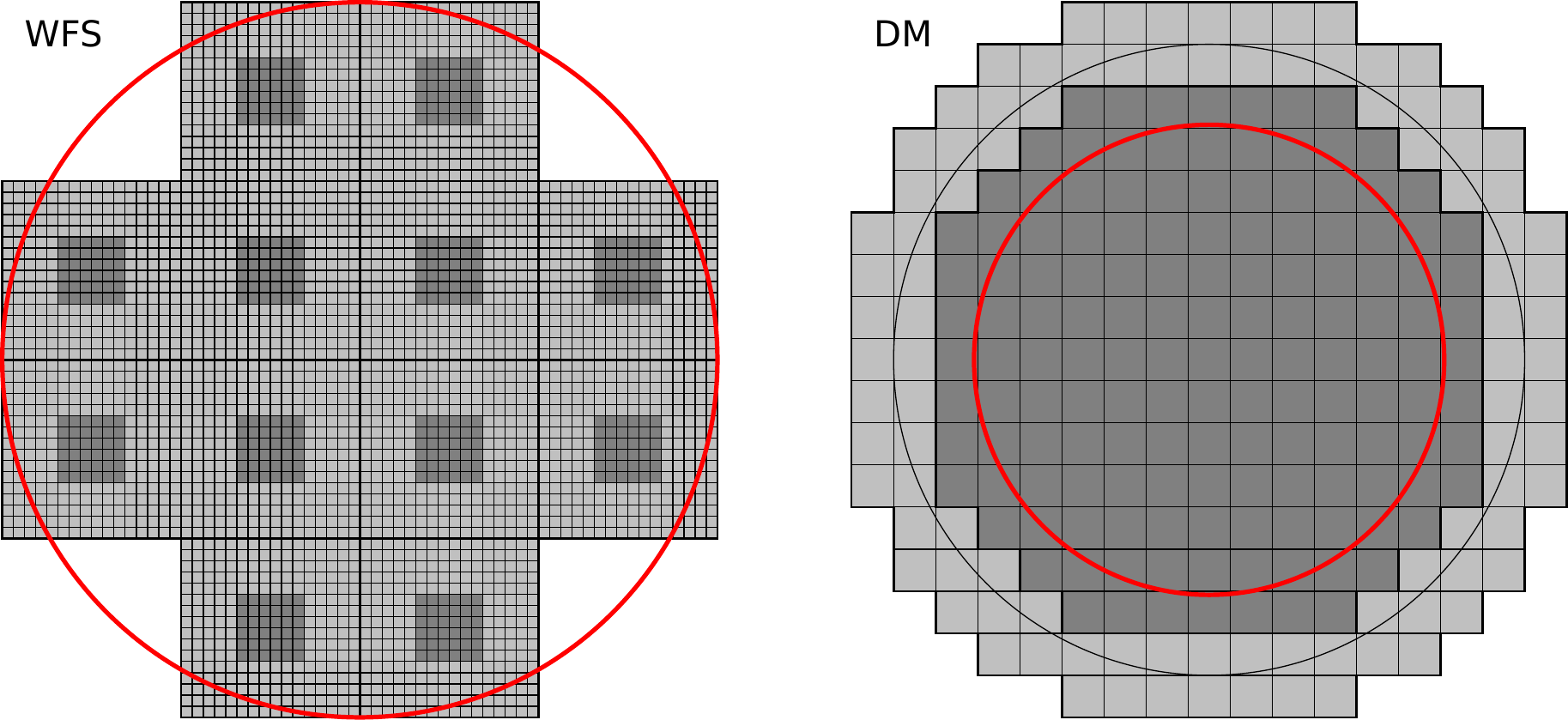}
    \caption{\textbf{Left (WFS)}: Geometry of the 12 wavefront sensor sub-apertures inside the pupil (red circle). Each sub-aperture (grey) is $16\times16$ pixels. Only the central $6\times6$ pixels (dark grey) are processed by the real-time controller. \textbf{Right (DM)}: Geometry of the deformable mirror at the M6 location. The \SI{28}{\milli\meter} diameter pupil (red circle) is smaller than the DM clear aperture (black circle). Out of the 241 available actuators (grey), $\sim145$ actuators participate in the wavefront correction (dark grey).}
    \label{Fig:WfsDmGeometry}
\end{figure}

\subsection{Wavefront sensor}\label{SSec:WavefrontSensor}

\begin{figure}
    \centering
    \begin{annotatedFigure}
        {\includegraphics[width=1.0\linewidth]{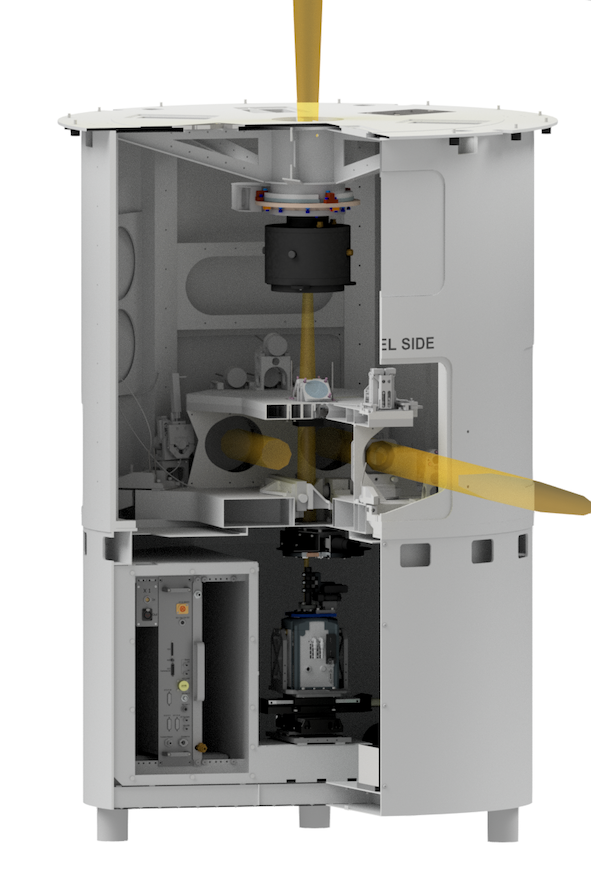}}
        \annotatedFigureBox{0.44,0.15}{0.64,0.21}{A}{0.64,0.18}
        \annotatedFigureBox{0.44,0.21}{0.64,0.31}{B}{0.64,0.26}
        \annotatedFigureBox{0.44,0.31}{0.64,0.37}{C}{0.64,0.34}
        \annotatedFigureBox{0.44,0.37}{0.64,0.42}{D}{0.64,0.395}
        \annotatedFigureBox{0.248,0.122}{0.36,0.358}{E}{0.248,0.122}
        \annotatedFigureBox{0.252,0.428}{0.759,0.636}{F}{0.252,0.636}
        \annotatedFigureBox{0.425,0.654}{0.625,0.774}{G}{0.425,0.774}
    \end{annotatedFigure}
    \caption{3D view of the dual feed ROS containing the NAOMI wavefront sensor located below the star separator. \textbf{A}: Field-patrolling XY table. \textbf{B}: Wavefront sensor camera. \textbf{C}: Pupil control unit. \textbf{D}: Relay lens and filter wheel. \textbf{E}: Wavefront sensor and motion control drive electronics. \textbf{F}: Star separator. \textbf{G}: De-rotator of the star separator.}
    \label{Fig:ViewRelayOpticsStructure}
\end{figure}
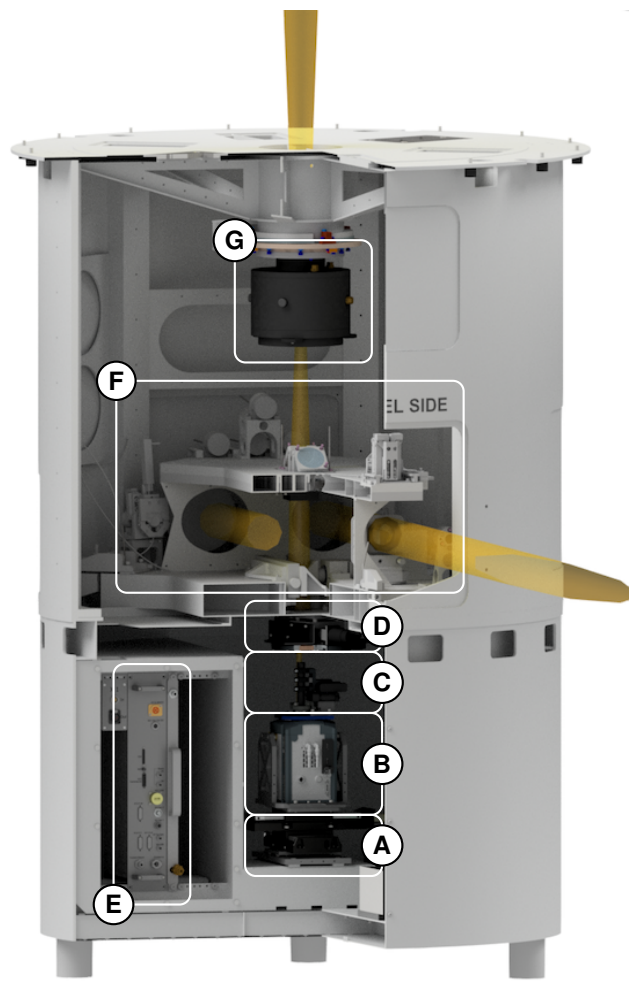

The wavefront sensor of NAOMI is located at the coudé focus of the AT, inside the so-called relay optics structure (ROS) that serves as an interface between the coudé train of the telescope and the light duct to the delay-line tunnel.
The upper part of the ROS contains a K-mirror-based field derotator and the star separator (STS) with the M9 dichroic; the lower part hosts an attenuation filter wheel and the wavefront sensor assembly mounted on a field-patrolling XY table.
Installed next to existing electronics for the STS, in a compartment of the lower ROS, a cooled electronics module drives both the pupil control unit (PCU) translation stage and the filter wheel, and serves as an interface between the WFS camera and the fibre to the SPARTA real-time controller.
An illustration of the ROS with the WFS inside is shown in Fig.~\ref{Fig:ViewRelayOpticsStructure}.

The wavefront sensor is a Shack-Hartman sensor (SHS).
It is based on an off-the-shelf ANDOR iXon Ultra 897 electron-multiplying CCD (EMCCD) camera paired with a $4\times4$ lenslet array with a \SI{512}{\micro\meter} sub-aperture pitch.
The camera is operated at a \SI{32}{\micro\meter} effective pixel size\footnote{The physical pixel size of the camera is \SI{16}{\micro\meter}. Being operated with a binning of 2, the effective pixel size is \SI{32}{\micro\meter}. Unless otherwise stated, effective pixels are always assumed herein.}, which gives $16\times16$ pixels per sub-aperture.
The camera plate scale of \SI{0.375}{\arcsec\per\pix} is close to the diffraction limit of \SI{0.3}{\arcsec} at R band (\SI{658}{\nano\meter}), and therefore introduces only a limited amount of non-linearities when it is operated with the diffraction-limited calibration source at the Nasmyth focus of the telescope.
The wavefront sensor geometry is illustrated in Fig.~\ref{Fig:WfsDmGeometry}.

The SHS is paired with a relay optic that scales the incoming pupil to the expected \SI{2.048}{\milli\meter} on the lenslet array.
A tele-centric lens, located after the M9 dichroic, makes the position of the pupil on the lenslet independent of the position of the sensor within the coudé field.
Because the pupil motion as a function of the derotator and azimuth angles can be as large as $17\%$ of the pupil ($68\%$ of a sub-aperture), the relay optic is mounted on a transverse translation stage to form the PCU, which can adjust the registration between DM and WFS.

The camera settings are adapted to the incoming flux level, as shown in Table~\ref{Tbl:ModesTable}.
As the incoming flux level decreases, the EM gain is first increased to reduce the effective readout noise, then the frequency is decreased.
In the most sensitive \SI{50}{\hertz} configuration, the latency-optimised "crop mode" is disabled to reduce the clock-induced charges \citep[CIC,][]{Hirsch+2013} by a factor of $\times4$.
To sustain this frequency, the horizontal shift (HS) speed needs to be increased again, which increases the readout noise, which in turn is finally compensated for by a further increase of the EM gain to $\times200$.
The vertical clock speed was set to \SI{0.5}{\micro\second}, the second fastest setting, to minimise both the frame transfer duration and the CIC.
To preserve charge transfer efficiency, an even faster setting of \SI{0.3}{\micro\second} would have required an increase of the vertical clock amplitude to a level that generates too much CIC.

The level of CIC shown in Table~\ref{Tbl:ModesTable} is more than an order of magnitude higher than the advertised \SI{0.002}{\electron/\pix} specification of the camera.
This is explained by two effects.
First, the \mbox{$2\times2$} binning increases the CIC by a factor of $\times4$.
Then, the crop mode collects a full detector width worth of CIC inside the width of the readout window, which amounts to another $\times3$ factor in our configuration.
Retrospectively, low-flux performance could have been improved further with an optical design that did not require a $2\times2$ binning.

\begin{table}
    \tiny
    \centering
    \begin{tabular}{cccccc}
        \hline
         Frequency  &   EM Gain   & Crop &     HS Speed     &      RON       &          CIC           \\
        \si{\hertz} &             &      & \si{\mega\hertz} & \si{\electron} & \si[per-mode=symbol]{\electron\per\pix} \\
        \hline
        \hline
            500     & $\times  1$ &   Y  &        10        &     64.8       &         0.036          \\
            500     & $\times 10$ &   Y  &        10        &      6.48      &         0.036          \\
            500     & $\times100$ &   Y  &        10        &      0.648     &         0.036          \\
            100     & $\times100$ &   Y  &         5        &      0.383     &         0.036          \\
             50     & $\times200$ &   N  &        10        &      0.324     &         0.012          \\
        \hline
    \end{tabular}
    \caption{WFS camera operation parameters listed in increasing sensitivity order. They include the frequency, the electron multiplication (EM) gain and the horizontal shift (HS) speed. Based on these parameters, the readout noise (RON) and clock-induced charges (CIC) performance improves, at the expense of saturation level first, as the EM gain increases, and latency then, as the frequency decreases.}
    \label{Tbl:ModesTable}
\end{table}

\subsection{Real-time controller}\label{SSec:RealTimeController}

\begin{figure*}
    \centering
    \includegraphics[width=0.9\linewidth]{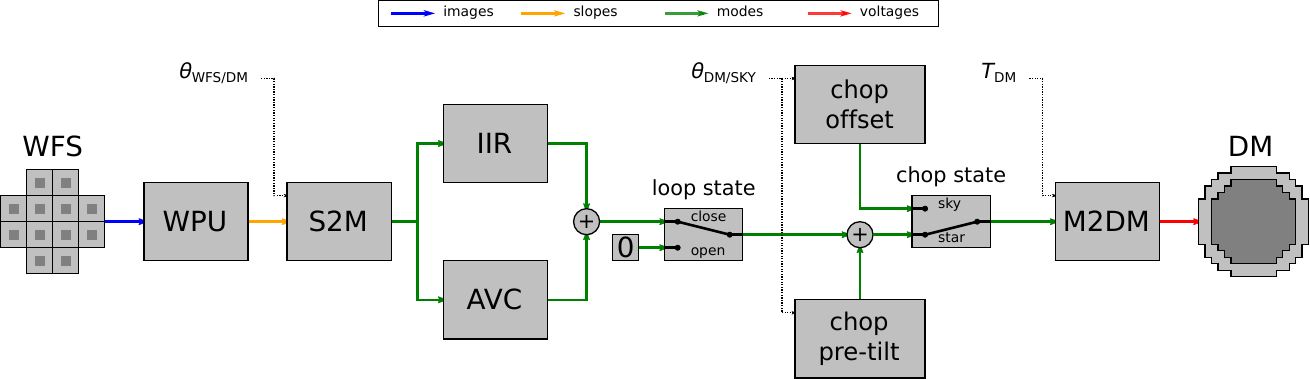}
    \caption{NAOMI control loop block diagram. The image stream from the WFS is converted into sub-aperture slope residuals by the WPU. The slope residuals are converted into mode residuals by matrix vector multiplication with the S2M matrix. Operating in parallel, the broad-band IIR and the AVC convert the mode residuals into a combined DM modal command when the control loop is closed. When chopping, a DM pre-tilt is added in the "star" state, and the DM is offset to in the "sky" state. The DM modal command is converted into actuator voltages by matrix vector multiplication with the M2DM matrix. The S2M matrix is rotated by $\theta_\mathrm{WFS/DM}$ to keep the modes aligned with the DM. The chopping offset and pre-tilt are rotated by $\theta_\mathrm{DM/SKY}$ to keep the chopping direction fixed on sky. To compensate for the temperature dependance of the DM, the M2DM matrix is scaled by the DM temperature $T_\mathrm{DM}$.}
    \label{Fig:BlockDiagram}
\end{figure*}

NAOMI employs the same SPARTA-light \citep{Suarez+2012} real-time controller as CIAO \citep{Scheithauer+2016}, the VLTI infrared wavefront sensor of the UTs.
SPARTA-light is a scaled-down version of the SPARTA platform \citep{Fedrigo+2006} deployed for the high-actuator-counts AO systems of Paranal.
The NAOMI RTC is based on four pipeline processing stages: a wavefront processing unit, a wavefront reconstruction unit, a control unit, and a DM conversion unit.
The wavefront processing unit receives pixels from the camera, extracts $64\times64$ images, removes a background, removes a threshold level, computes the intensity per sub-aperture, applies a weighting map, and computes a weighted centre of gravity for each of the 12 sub-apertures.
To keep the latency at a minimum, only the central $6\times6$ pixels are processed by the RTC.
The calculated intensity per sub-aperture, and its average over the sub-apertures, are used for acquisition (see Sect.~\ref{SSec:AcquisitionProcedure}) and target magnitude estimation.
From the slopes of the 12 sub-apertures, the wavefront reconstruction is performed by matrix vector multiplication (MVM) with the "slope-to-mode" (S2M) control matrix, delivering a result in a modal space of 14 first \si{\micro\meter\rms}-normalised Zernikes excluding piston.
The control is based on two pipelines running in parallel: an infinite impulse response (IIR) and an anti-vibration control (AVC).
The conversion from Zernike-space to actuator-space is performed by MVM with a mode-to-DM (M2DM) matrix, which is the result of the calibrated influence functions and scaled by the DM gain measured on sky (see Sec.~\ref{SSec:DeformableMirror}).
The newly calculated DM commands are finally converted into a series of eight commands, interpolated from the previous commands, in order to reduce the excitation of the DM resonance.
The commands are finally converted into the 14 bits unsigned integer format accepted by the DM.
The control loop is summarised in the flowchart of Fig.~\ref{Fig:BlockDiagram}.

\subsection{Acquisition procedure}\label{SSec:AcquisitionProcedure}

The acquisition procedure for the ATs with NAOMI can be decomposed into three individual steps before closing the AO loop.
These steps are identical to those of the former STRAP system, and of the MACAO and CIAO systems on the UTs.

First, a new preset is sent from one of the instruments, through the Interferometer Supervisory Software, to the AT.
Relevant to NAOMI, the preset contains the on-axis coordinates of the telescope, the coordinates, and the R magnitude of the AO guide star.
The difference between the two coordinates drives the position of the XY table and therefore the off-axis offset of the wavefront sensor at the coudé focus.
The R magnitude sets an optimised rate for the WFS camera operating without EMCCD amplification, and possibly triggers the insertion of a filter to reduce the flux on very bright objects.
This so-called detection configuration is chosen to optimally detect the target without saturating, in a \SI{6x6}{\arcsec} field of view of the WFS, within a \SI{1}{\second} long accumulation of camera frames.

Second, when the telescope follows a blind trajectory, the target is detected in order to align the field and the photometric pupil on the WFS.
To do so, a \SI{1}{\second} long accumulation of background-corrected WFS frames is measured to increase the S/N.
This dataset is processed in two different ways.
The 12 sub-apertures are averaged and stacked into a single \SI{6}{\arcsec}$\times$\SI{6}{\arcsec} image of the field in order to measure the target offset from the centre, converted into a telescope offset.
Then, the averaged fluxes in the four corner sub-apertures are used to measure the photometric pupil alignment, and correct it with the PCU.
The flux of the target is also used to correct possible mistakes in the user-provided R magnitude.
If the star cannot be detected within that field of view (due to an error in the pointing model of the telescope or in the coordinates of the guide star), a disco-ball-looking shape can be applied on the DM to point each sub-aperture to a different region of the sky, in a mosaic fashion, which turns the SHS into a \SI{24x24}{\arcsec} imager.

Third, this measured target flux is used to optimise the full parameters of the AO system.
The neutral density filter, camera rate, EMCCD gain, loop gain, and number of controlled modes are adjusted based on a performance-optimal prescription that we illustrate in Fig.~\ref{Fig:PerformanceVsFlux}.
The flux of the acquired target is checked against the optimisation prescription once more, to guarantee the best operation configuration for the flux incident on the WFS, as measured by the AO system itself.
The AO loop is briefly closed to measure the amount of defocus present on the DM.
To maximise the DM stroke, this defocus is compensated for by adjusting the longitudinal position of the secondary mirror of the telescope.
As long as the target flux is high enough, the gain of the DM is also measured on sky as part of this step, with the coma modulation technique described in Sec.~\ref{SSec:DeformableMirror}.

\begin{figure}
    \centering
    \includegraphics[width=0.8\linewidth]{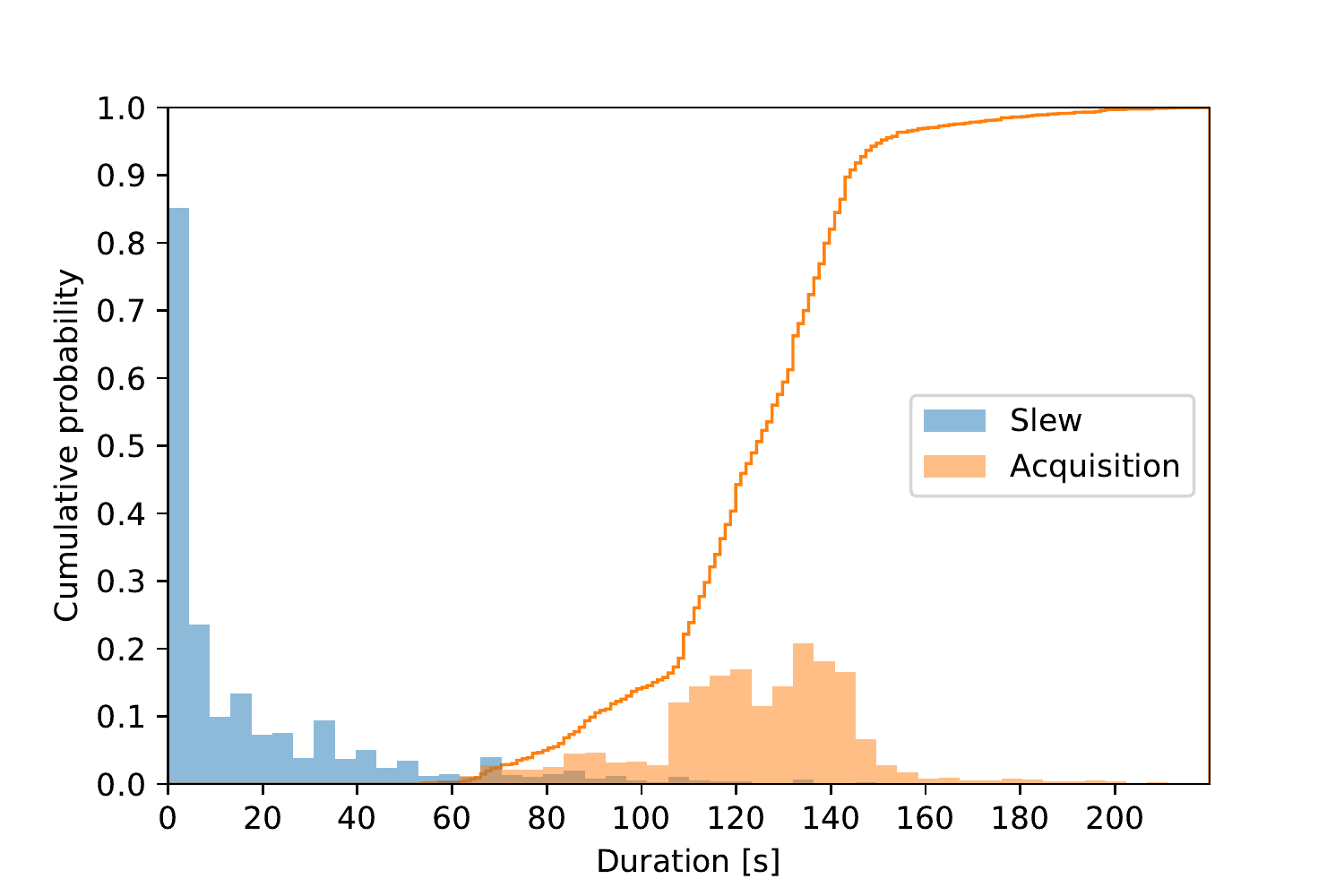}
    \caption{Histogram of the slew and acquisition durations of the auxiliary telescopes with NAOMI. The slew duration is dominated by the short \SI{10}{\second} long switches between targets and calibrators. The acquisition duration is below \SI{150}{\second} in $90\%$ of the time.}
    \label{Fig:AcquisitionTiming}
\end{figure}

When these three steps are completed, the AO loop is ready to be closed.
This entire acquisition sequence has been fully automated to reduce the workload on the operators and keep the acquisition duration to a minimum (see Fig.~\ref{Fig:AcquisitionTiming}).
The automatic acquisition works well when there are no other bright targets around the guide star.
For more complex fields, the operator has the possibility to manually execute the acquisition sequence and specifically select the correct guide star.
For narrow equal-brightness binaries, the operator may even have to adjust the FoV over which the AO loop operates (from $6\times6$ to $16\times16$ pixels per sub-aperture), in order for the SHS to measure reliable slopes from either one or both targets, depending on the separation.

\subsection{Pupil registration}\label{SSec:PupilRegistration}

When the AO loop is closed, the pupil registration loop is started.
The pupil misregistration error signal is calculated from the flux imbalance between the side sub-apertures.
In contrast to the direct method, which correlates the DM increments to the wavefront residuals \citep[e.g.][]{Bechet+2012}, this method assumes that the photometric pupil is a good proxy for the DM.
This is the case on the ATs, where most of the misalignment errors are introduced between the DM and the WFS, by the azimuth axis and the field derotator.
A control loop sends corrections once every 10 seconds, not only to the PCU to re-align the DM to the WFS, but also to the XY table to compensate for the tip-tilt that a PCU motion alone would introduce.
Due to this effect on the tip-tilt, and because that PCU and XY table motions are not synchronised, the correction commands are clipped to induce no more than a tenth of a K-band PSF equivalent per pupil registration loop cycle.
The control bandwidth being very slow, it relies on the pre-alignment that is carried out as part of the acquisition procedure, to rapidly reach an alignment state better than $10\%$ of a sub-aperture size.

\subsection{Chopping}\label{SSec:Chopping}

Even though not ideal in terms of instrumental thermal background rejection, chopping is performed directly with the deformable mirror of NAOMI.
This is unchanged from when STRAP was installed and the M6 tip-tilt mirror was the chopping element.
The \mbox{ALPAO} DM241 is a deformable mirror with a comparably large tip-tilt stroke, but still not sufficient to chop with an amplitude many times the \SI{1.15}{\arcsec} telescope lobe size in the thermal infrared (\SI{10}{\micro\meter}).
To increase the chopping amplitude, the telescope is tilted towards the mid-point between the on-target and on-sky positions.
This lets the DM chop from two extreme tilt positions, increasing the stroke by a factor $\sim2$.
This balanced chopping on the DM has the additional benefit of keeping the temporal average DM position tip-tilt-free, which mitigates the creep effect described in Sec.~\ref{SSec:DeformableMirror}.
Like all other chopping elements of VLTI, the timing of the chopping sequence is derived from the absolute observatory time, which allows precise synchronisation with the VLTI tip-tilt sensor IRIS, and the mid-infrared MATISSE instrument.
Once the chopping sequence is activated, the AO loop alternates between loop closed for the on-target phase, and loop opened with a flat DM chopped away for the on-sky phase.
To prevent DM saturations even further, the temporally averaged tip-tilt content of the DM, without the chopping offsets, is offloaded to the main axes of the telescope.

\section{Adaptive optics performance}\label{Sec:Performance}

Before shipment to Paranal, the functionality and performance of the NAOMI system were verified and optimised on a test bench that was developed in Garching for the occasion.
A horizontal optical bench, containing the DM assembly in the same attitude as on the AT and a calibration source, was mounted on top of a replica of the ROS containing the K-mirror, the M9 dichroic feeding the actual WFS assembly, and a so-called Strehl J-band camera.
In this section we present the stand-alone performance achieved by the NAOMI system as measured on the Garching test bench and as verified on sky in Paranal.

\subsection{Transfer function}\label{SSec:TransferFunction}

\begin{figure}
    \centering
    \includegraphics[width=0.8\linewidth]{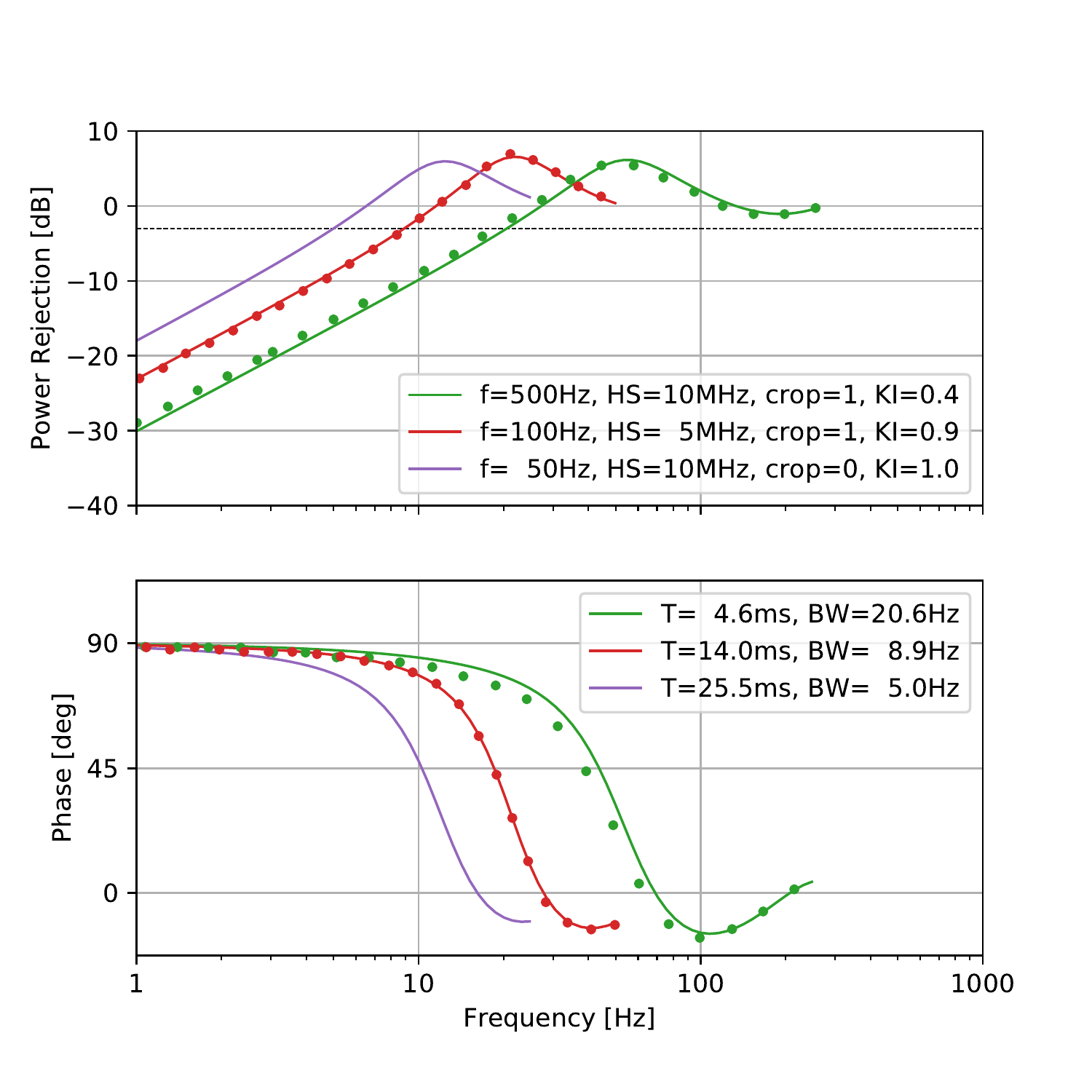}
    \caption{Measurements (dots) and models (lines) for the wavefront transfer function for the three loop frequency and gain settings adopted for the NAOMI system. The settings are defined by the loop frequency (f), the HS speed, the crop (crop=1) or region of interest (crop=0) mode, and the integrator gain (KI). The total delay (T) and the \SI{-3}{\deci\bel} bandwidth (BW) of the models are also given. Measurements and models are in agreement, confirming that the delays and scaling of the control loop are properly accounted for. See Sect.~\ref{SSec:TransferFunction} for details.}
    \label{Fig:TransferFunction}
\end{figure}

\begin{figure}
    \centering
    \includegraphics[width=0.95\linewidth]{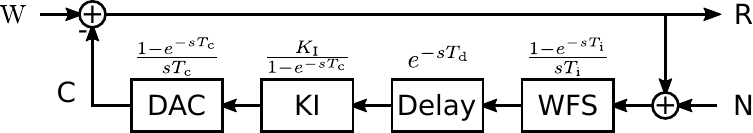}
    \caption{NAOMI control loop model. The incident wavefront W is compensated for by the command C applied on the DM. The residual wavefront R is averaged over the integration time $T_\mathrm{i}$ by the WFS, delayed by $T_\mathrm{d}$, integrated by a controller with a gain $K_\mathrm{I}$, and converted into commands C by an digital-to-analog converter (DAC). The cycle time $T_\mathrm{c}$ is the inverse of the loop frequency. The pure delay $T_\mathrm{d}$ includes contributions from the camera frame transfer and readout time, the RTC computation time, and the DM response time. The Laplace transform of each idealised operation is shown above its respective block.}
    \label{Fig:ControlLoop}
\end{figure}

The wavefront transfer function of the system was measured on the test-bench to verify its behaviour.
The results are shown in Fig.~\ref{Fig:TransferFunction} and compared to a model illustrated in Fig.~\ref{Fig:ControlLoop}.
At the \SI{500}{\hertz}, \SI{100}{\hertz}, and \SI{50}{\hertz} settings, the \SI{-3}{\deci\bel} bandwidth is measured at \SI{18.6}{\hertz}, \SI{8.9}{\hertz}, and \SI{5.0}{\hertz} respectively.

The transfer function model also verifies the total delay $T$ introduced in the control loop.
It is the sum of half the integration time $T_\mathrm{i}/2$ from the detector sliding average, a pure delay $T_\mathrm{d}$, and half the cycle period $T_\mathrm{c}/2$ from the DM digital-to-analog converter zero-order hold.
The pure delay $T_\mathrm{d}$ includes the camera frame transfer time $T_\mathrm{FT}$ and readout time $T_\mathrm{RO}$, the controller computation time $T_\mathrm{RTC}$, and the DM response delay $T_\mathrm{DM}$.
The readout time $T_\mathrm{RO}$ depends on the camera HS speed and crop mode, which affects the readout noise.
Taking into account the relationship $T_\mathrm{c}=T_\mathrm{i}+T_\mathrm{FT}$ between the cycle, integration, and readout times, the total delay reads
\begin{equation}
    \arraycolsep=1.4pt
    \begin{array}[t]{cccccccccccl}
        \underbrace{T_{}}
            &=& \underbrace{T_\mathrm{c}}
                &+& \underbrace{T_\mathrm{FT}/2}
                    &+& \underbrace{T_\mathrm{RO}}
                        &+& \underbrace{T_\mathrm{RTC}}
                            &+& \underbrace{T_\mathrm{DM}},
                                & \\
        \scriptstyle\SI{4.6}{\milli\second}
            && \scriptstyle\SI{2}{\milli\second}
                &&
                    && \scriptstyle\SI{1.54}{\milli\second}
                        &&
                            &&
                                & \scriptstyle\ @\SI{500}{\hertz} \\
        \scriptstyle\SI{14.0}{\milli\second}
            && \scriptstyle\SI{10}{\milli\second}
                && \scriptstyle\frac{\SI{0.27}{\milli\second}}{2}
                    && \scriptstyle\SI{2.93}{\milli\second}
                        && \scriptstyle\SI{0.35}{\milli\second}
                            && \scriptstyle\SI{0.6}{\milli\second}
                                & \scriptstyle\ @\SI{100}{\hertz} \\
        \scriptstyle\SI{25.5}{\milli\second}
            && \scriptstyle\SI{20}{\milli\second}
                &&
                    && \scriptstyle\SI{4.38}{\milli\second}
                        &&
                            &&
                                & \scriptstyle\ @\SI{50}{\hertz}\\
    \end{array}
,\end{equation}
which is annotated with the independently measured delay contributions at the three loop frequency settings.
The agreement between transfer function measurements and models, illustrated in Fig.~\ref{Fig:TransferFunction}, confirms that the delay and scaling of the control loop are properly accounted for.

\subsection{Performance versus magnitude}\label{SSec:PerformanceVsMagnitude}

\begin{figure}
    \centering
    \includegraphics[width=0.8\linewidth]{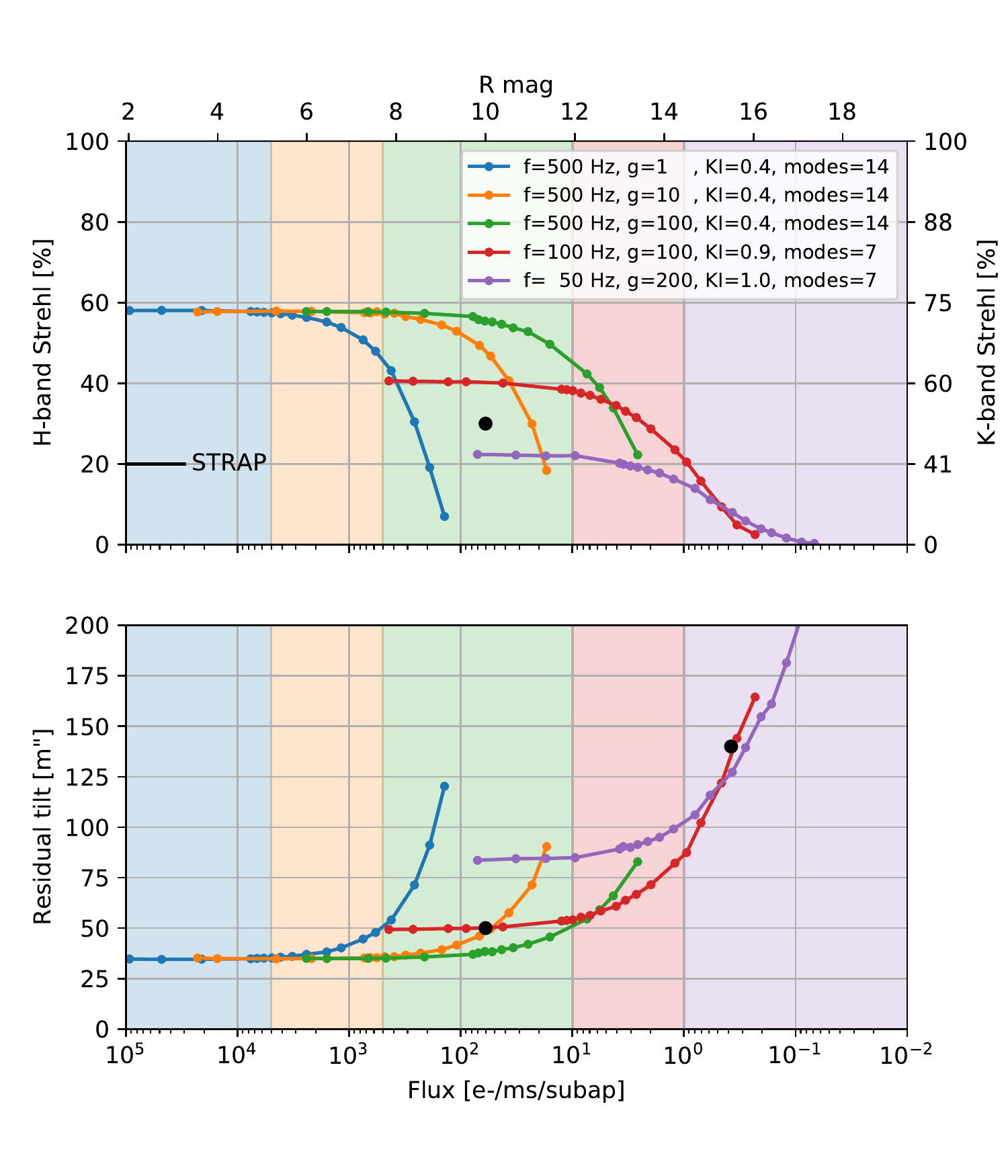}
    \caption{Laboratory-measured Strehl (\textbf{top}) and residual tilt (\textbf{bottom}) performance of NAOMI versus WFS incident flux, at 80\% seeing conditions, i.e. $\lambda/r_0=\SI{1.1}{\arcsec}$ and $\tau_0=\SI{2.5}{\milli\second}$ at $\lambda=\SI{500}{\nano\meter}$.
    The former performance level of STRAP is also indicated.
    The coloured lines correspond to the performance measurements for each loop frequency (f), EM gain (g), loop gain (KI), and controlled mode (modes) settings.
    The coloured areas are the optimal settings for a given flux range.
    The WFS incident flux is converted into an R magnitude assuming a transmission to the WFS of 10\%.
    The specifications for the Strehl and residual tilt vs. R magnitude are shown as black dots.
    The transitions between modes at \SI{500}{\hertz} were chosen based on saturation limits.
    The transition between modes at lower frequencies were chosen at a flux level two times higher than the Strehl performance intersections, to help with the stability of the AO loop close these transitions.}
    \label{Fig:PerformanceVsFlux}
\end{figure}

The H-band Strehl and residual tip-tilt as a function of the incident flux was initially characterised on the test bench.
This was done by injecting a turbulent perturbation on the DM while closing the loop on the calibration source.
Due to the high DM actuator count, the contribution of the DM fitting error\footnote{In median \SI{1.1}{\arcsec} seeing conditions, the DM fitting error in H band is a negligible $\sigma^2_\mathrm{DM} = \SI{0.05}{\radian^2}$.} could be properly simulated.
Both Strehl and residual tip-tilt were estimated from the residual motion of the DM with the AO loop closed.
The test bench results shown in Fig.~\ref{Fig:PerformanceVsFlux} are in agreement with the performance achieved on sky.

\subsection{Piston control}\label{SSec:PistonControl}

\begin{figure}
    \centering
    \includegraphics[width=0.8\linewidth]{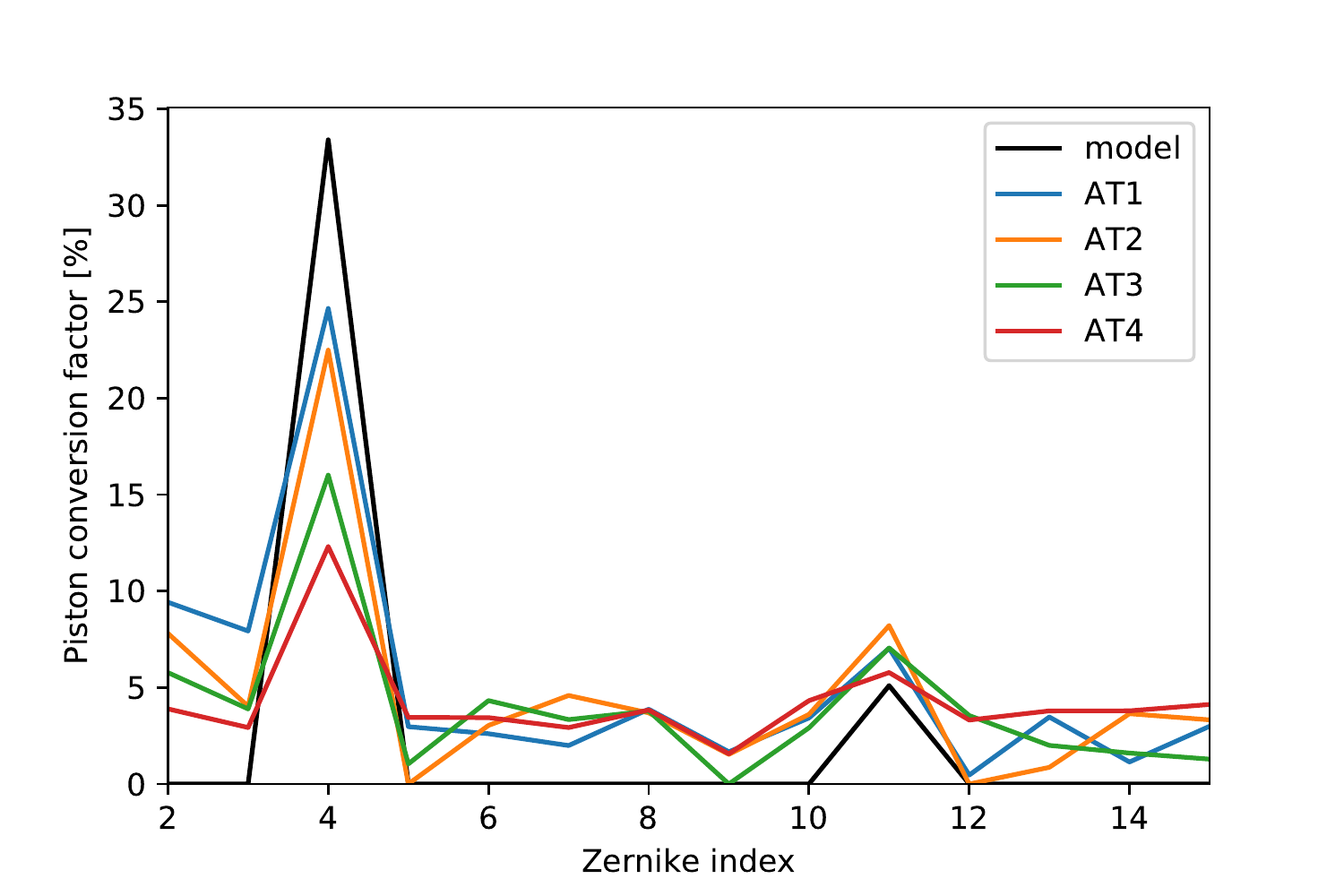}
    \caption{Piston conversion factor for each mode of each AO system, measured by injecting a modulation on the DM and detecting it with the GRAVITY fringe tracker. The conversion factor is the ratio between the piston measured in $\si{\micro\meter\rms}$ and the modal modulation in $\si{\micro\meter\rms}$. A model with optimal single-mode apodisation confirms the high conversion factors for radial-only focus and spherical modes; the amplitude mismatch arises because the DM gain (see Sec.~\ref{SSec:DeformableMirror}) was not yet corrected at the time of the measurement. The small conversion factor for tip-tilt is compatible with a small $2.5\%$ photometric pupil shift. All other modes show little conversion factor, probably within the noise of the measurement.}
    \label{Fig:PistonFree}
\end{figure}

The piston-free property of the controlled modes could only be verified on sky after deployment of the NAOMI systems on the interferometer.
A single frequency modulation was injected on each mode of each AO system.
The frequency was chosen to be high enough to avoid residual turbulence contamination, yet low enough to be properly sampled at the AO loop frequency.
The amount of piston introduced was estimated with the GRAVITY fringe tracker.
The results are illustrated in Fig.~\ref{Fig:PistonFree}.
The presence of piston on the radial-only modes, focus (Z4), and spherical (Z11) is compatible with a pupil apodisation by the GRAVITY fringe tracker single-mode fibre.
The slightly larger piston on tip (Z2) and tilt (Z3) is compatible with a small $\sim2.5\%$ shift of the effective photometric pupil on the DM.
Technically, it should be possible to change the piston-free definition of the NAOMI modes to include the apodisation effect, but the modes would then have to adapt to the instrument in use (GRAVITY and PIONIER apodise, but MATISSE does not).
This observation of apodisation-induced piston should serve as a reminder that a good AO correction is a requirement to co-phasing two instruments with different interferometric lobes\footnote{See \citet{Mege2002} or \citet{Perrin&Woillez2019} for a definition of the single-mode interferometric lobe; a similar definition exists for non-single-mode instruments but depends on the beam combination geometry.}.

\subsection{Non-common path aberrations}\label{SSec:NCPA}

\begin{figure}
    \centering
    \includegraphics[width=0.8\linewidth]{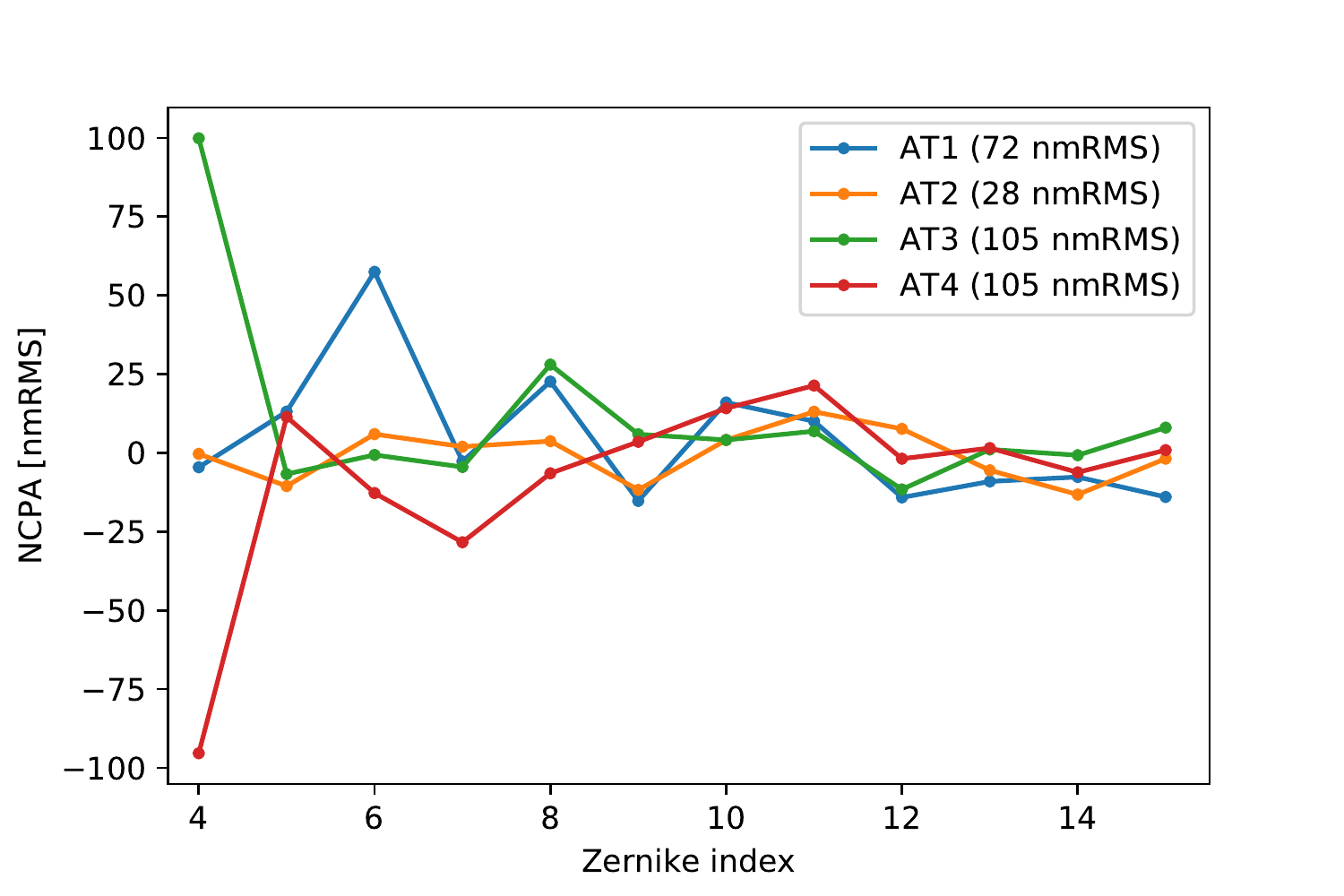}
    \caption{Non-common path aberrations measured between NAOMI and the IRIS tip-tilt sensor of the VLTI.}
    \label{Fig:NCPA}
\end{figure}

Non-common path aberrations (NCPA) were measured all the way to the IRIS tip-tilt sensor of VLTI \citep{Gitton+2004}, using the telescope Nasmyth beacon in day time.
Sequentially on each controlled mode starting with defocus, a modulation is injected into the closed AO loop at the level of the WFS slope measurements.
The flux of the brightest pixel is recorded on IRIS in closed loop.
The modulation on each mode contains a high \SI{25}{\hertz} frequency and two periods of a low \SI{0.25}{\hertz} frequency.
The amplitude of the \SI{25}{\hertz} detected on IRIS reaches a minimum when the slow period crosses the NCPA offset.
Two periods give at least three consecutive minima and an easy measurement mechanism from their temporal spacing.
As illustrated in Fig.~\ref{Fig:NCPA}, differential aberrations are below $\sim\SI{100}{\nano\meter}$, confirming the quality of the WFS and the VLTI beam train.

\section{Impact on the VLTI}\label{Sec:ImpactOnVlti}

This section explores the impact of NAOMI on the VLTI instruments.
It makes extensive use of the QC1 database, which is publicly available on the ESO website at \url{http://www.eso.org/qc}, and contains the output of the data reduction pipelines of the instruments.
The dataset for the former STRAP tip-tilt system covers the period 2018 January 1 to 2018 September 6, before the VLTI shutdown.
The dataset for NAOMI starts at the end of the first commissioning on 2018 November 17, and avoids the second commissioning between 2019 February 24 and 2019\ March 5 where the NAOMI systems may have been used, part of the time, in non-standard configurations for dedicated tests.

\subsection{Strehl stability and fringe tracking}\label{SSec:StrehlStability}

\begin{figure}
    \centering
    \includegraphics[width=0.8\linewidth]{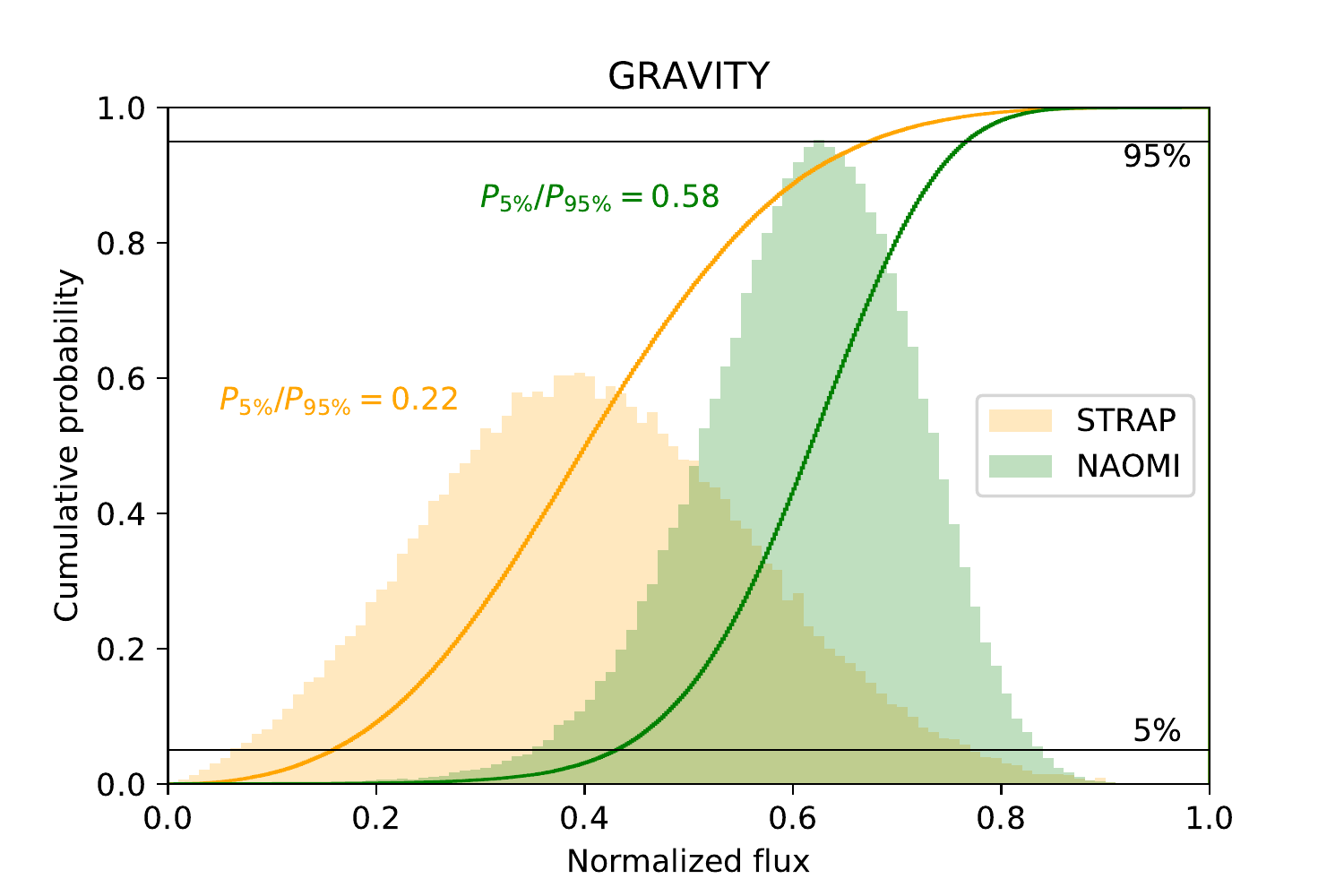}
    \caption{Typical instantaneous injected flux histograms for the GRAVITY fringe tracker for STRAP and NAOMI, in median seeing conditions. The $P_{5\%}/P_{95\%}$ metric, ratio of the injected fluxes at $5\%$ and $95\%$, is derived from these histograms. STRAP at $P_{5\%}/P_{95\%}=0.22$ has significantly more events at low flux than NAOMI at $P_{5\%}/P_{95\%}=0.58$.}
    \label{Fig:FluxHistogram}
\end{figure}

The effect of adaptive optics correction order on the instantaneous single-mode coupling was predicted by \citet[][see Fig.~6 therein]{Tatulli+2010}, including a specific application to the ATs of the VLTI.
Fig.~\ref{Fig:FluxHistogram} finds a median seeing condition instantaneous Strehl measured by the GRAVITY fringe tracker in qualitative agreement with these predictions.
Tip-tilt only correction is not optimal for fringe tracking because of significant flux dropouts.

\begin{figure}
    \centering
    \includegraphics[width=0.8\linewidth]{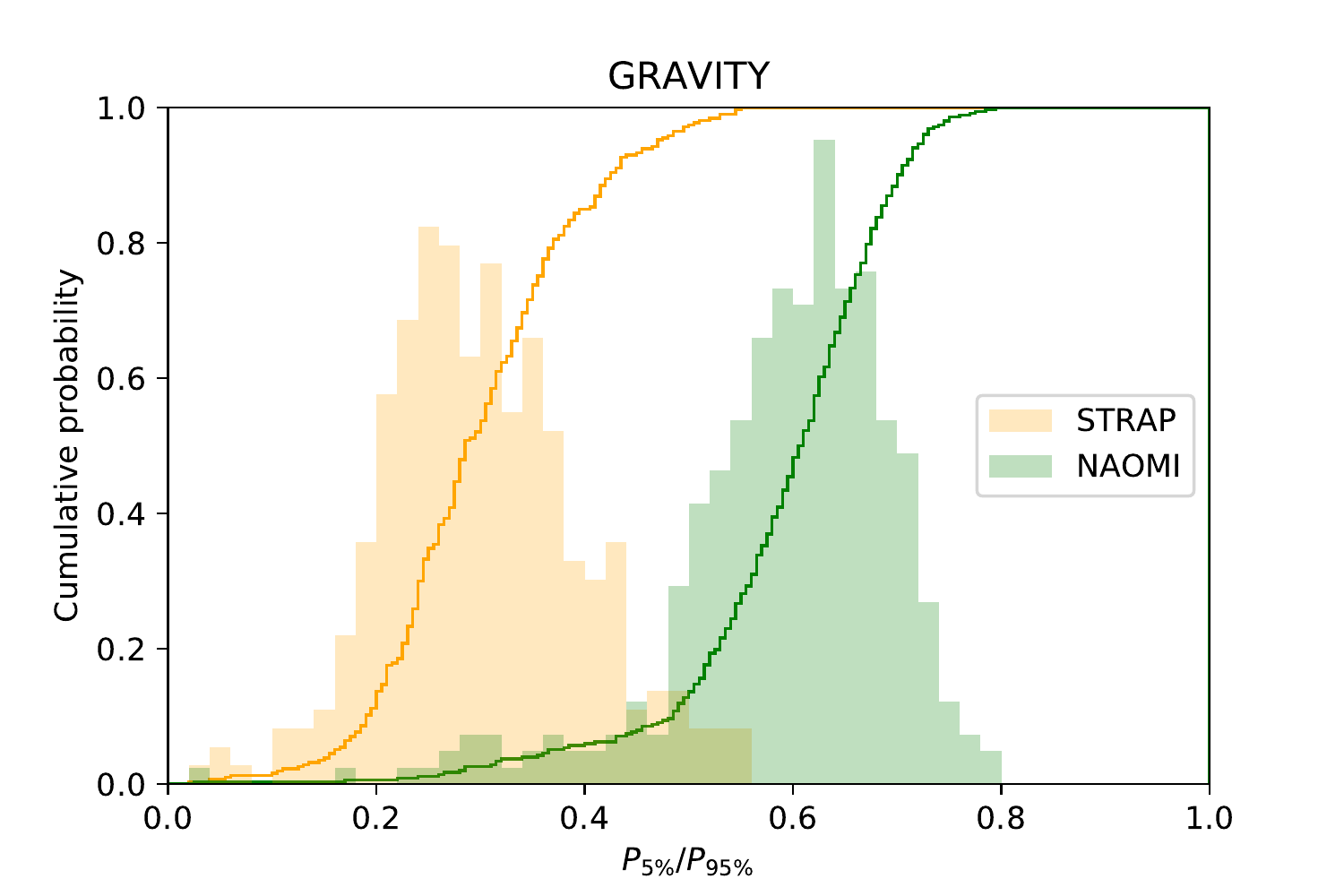}
    \caption{Comparison between STRAP tip-tilt and NAOMI AO of the $P_{5\%}/P_{95\%}$ injection metric histograms for the GRAVITY fringe tracker. The flux dropouts, represented by $P_{5\%}/P_{95\%}$ values close to zero, are significantly reduced by NAOMI.}
    \label{Fig:P05P95_GRAVITY}
\end{figure}

\begin{figure}
    \centering
    \includegraphics[width=0.8\linewidth]{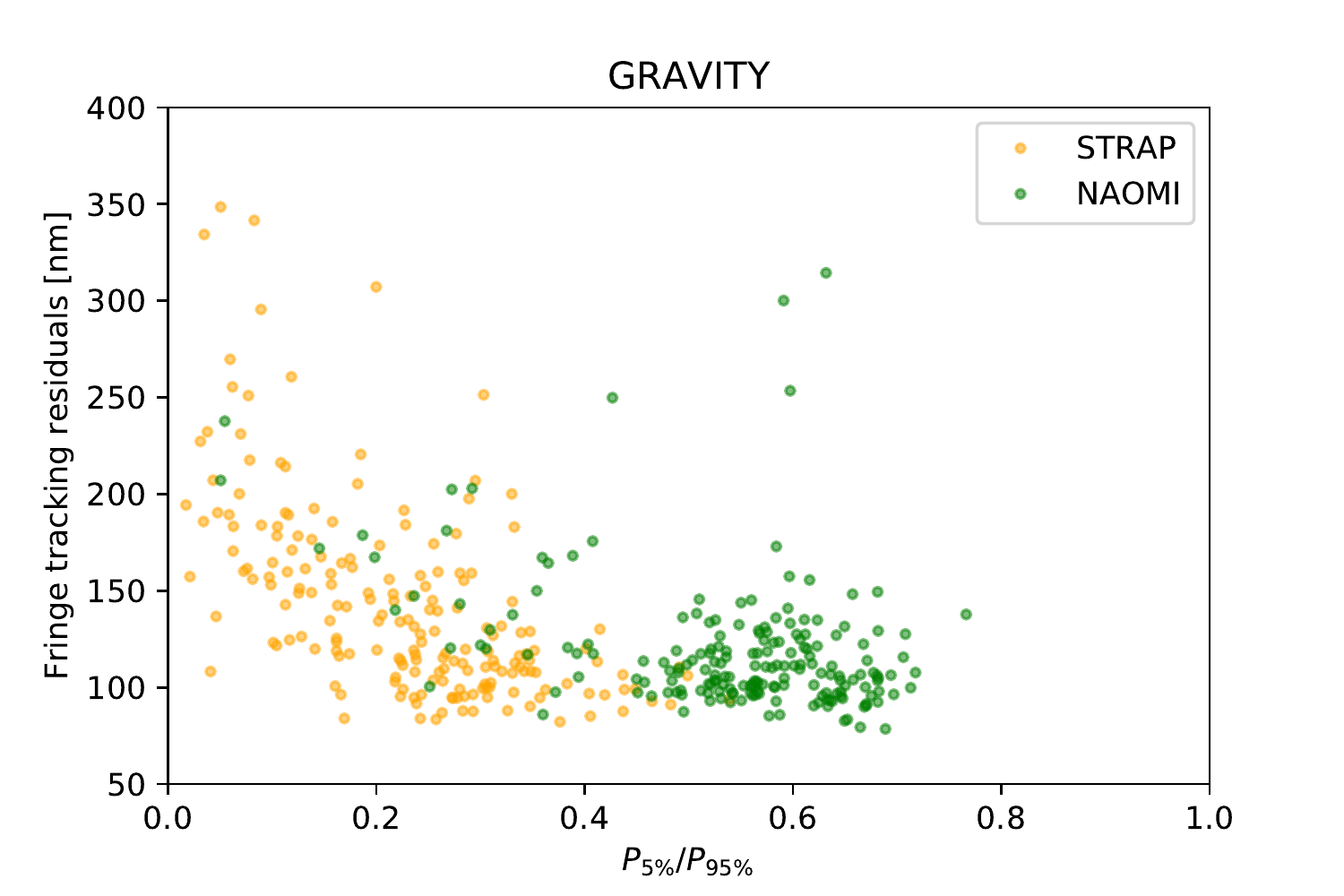}
    \caption{Correlation between the GRAVITY fringe tracker residuals and the $P_{5\%}/P_{95\%}$ injection stability metric, comparing STRAP and NAOMI. With NAOMI, the fringe tracking residuals tend to stay in the \SIrange{80}{150}{\nano\meter} range because the $P_{5\%}/P_{95\%}>0.5$ injection stability is high. This figure does not include $20\%$ of the open-dome nights with wind speeds higher than $\SI{9}{\meter\per\second}$, where telescope vibrations affect fringe tracking residuals \citep{Woillez+2016}.}
    \label{Fig:FringeTrackingResiduals}
\end{figure}

To explore the statistics of this improvement, we designed a dedicated metric noted $P_{5\%}/P_{95\%}$.
It represents the ratio of the flux at low $5\%$ injection to the flux at high $95\%$ injection.
This ratio tends to zero in presence of flux dropouts, and tends to one when the injection is perfectly stable.
Fig.~\ref{Fig:P05P95_GRAVITY} shows a comparison of the GRAVITY fringe tracker injection stability between the former tip-tilt system and the new AO.
It confirms a significant reduction in flux dropouts.
As a consequence,  Fig.~\ref{Fig:FringeTrackingResiduals} shows the support role that AO plays in reducing fringe tracking residuals.
With STRAP, the metric would very often drop below $0.2$ and the fringe tracking residuals increase.
When the $P_{5\%}/P_{95\%}$ metric is above $0.5$, as is the case most of the time with NAOMI, fringe tracking residuals tend to stay within the \SIrange{80}{150}{\nano\meter} range.

\subsection{Transmission}\label{SSec:Transmission}

\begin{figure}
    \centering
    \includegraphics[width=0.8\linewidth]{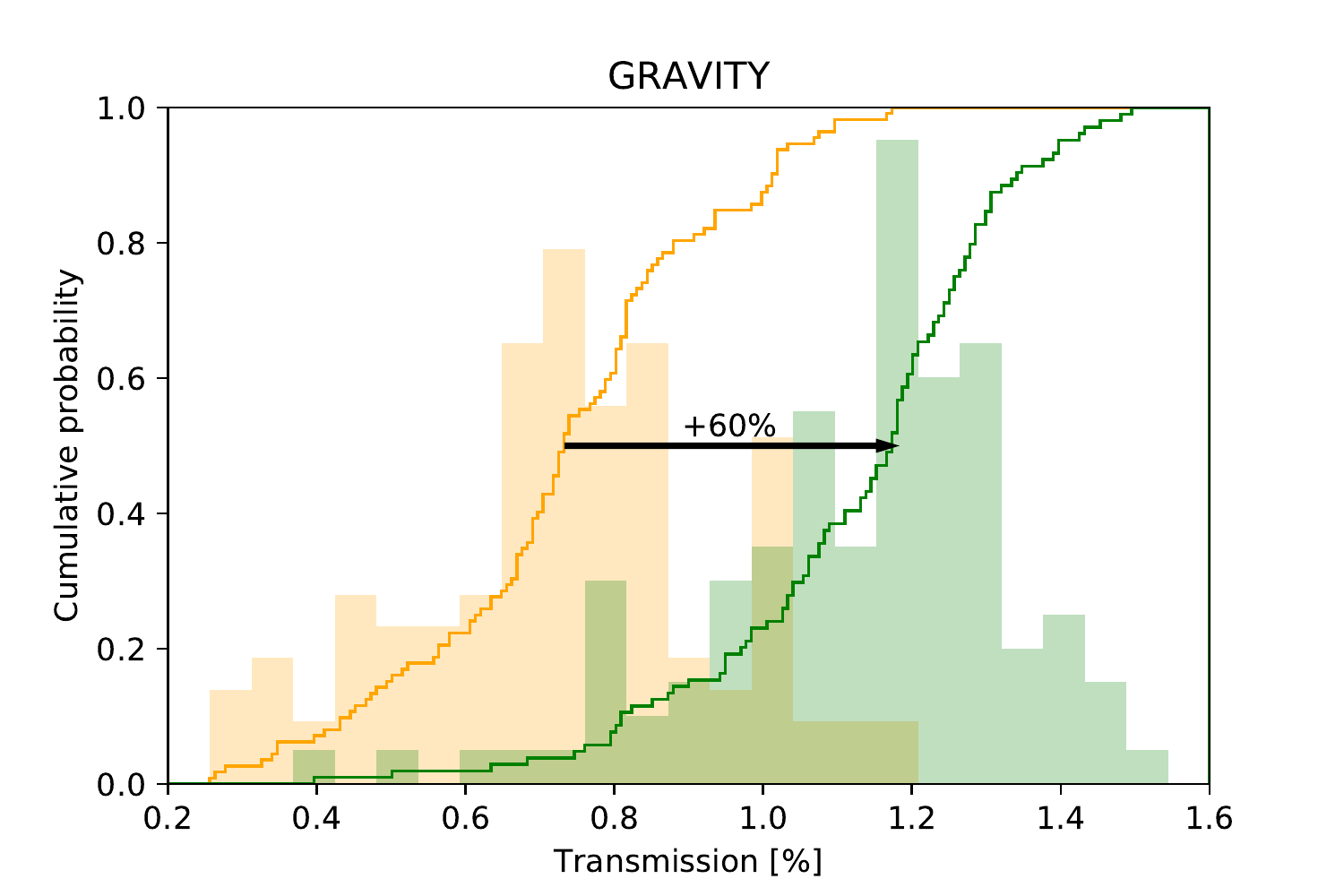}
    \includegraphics[width=0.8\linewidth]{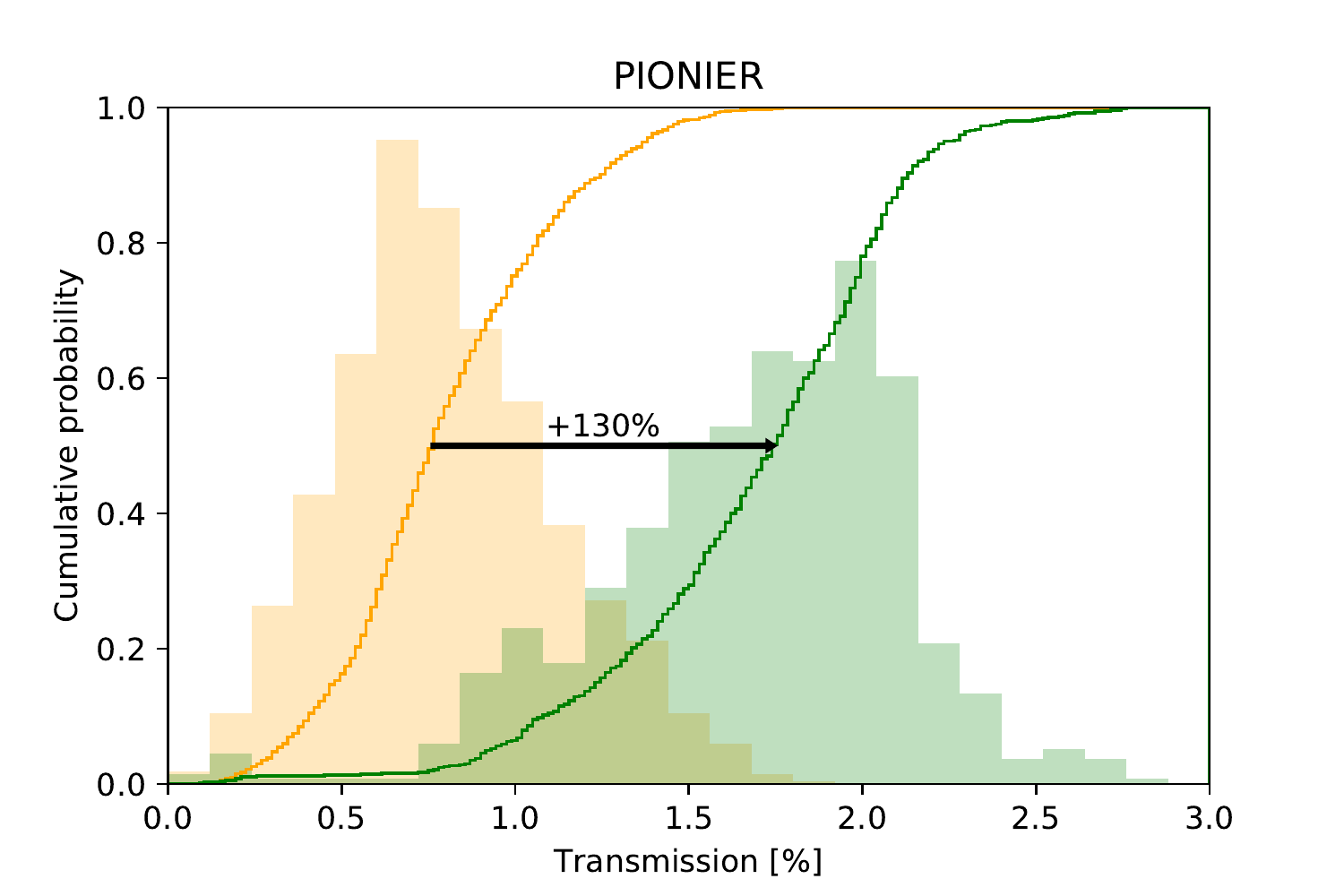}
    \caption{Comparison of GRAVITY and PIONIER transmission between STRAP and NAOMI. NAOMI shows a transmission improvement of $+60\%$ for GRAVITY and $+130\%$ for PIONIER.}
    \label{Fig:TransmissionGravity}
\end{figure}

The deployment of NAOMI also improves the average injection efficiency into the single-mode instruments GRAVITY (K band) and PIONIER (H band) by $+60\%$ and $+130\%,$ respectively, as demonstrated in Fig.~\ref{Fig:TransmissionGravity}, on the condition that the R-band magnitude remains in the high Strehl regime ($R<\SI{12}{\mag}$).
This is a direct consequence of the higher Strehl ratio delivered by the AO system, and the resulting higher single-mode coupling efficiency.
The improvement level, however, is lower than the predictions given in Fig.~\ref{Fig:PerformanceVsFlux} for an effective seeing of \SI{1.1}{\arcsec}: +70\% in K band and +190\% in H band.
This may be explained by a better-than-expected median effective seeing of \SI{0.9}{\arcsec}.
The transmission improvement is larger at shorter wavelengths, and invites an extension of the VLTI instrument suite to the J band.
We also investigated the effect of the improved Strehl ratio on the interferometric observable stability and error bars, but did not observe any second-order improvement beyond what the improved transmission brings.

The limiting magnitudes of the PIONIER and GRAVITY instruments are increased by $+\SI{1}{\mag}$ under the high Strehl regime requirement of $R<\SI{12}{\mag}$, to $K=\SI{9}{\mag}$ for GRAVITY on-axis and $H=\SI{9}{\mag}$ for PIONIER without dispersion.
For PIONIER, we see a direct consequence of the improved transmission.
For GRAVITY, half of the improvement is related to transmission, and the other half is due to the Strehl stabilisation.
With a more stable Strehl ratio, the GRAVITY fringe tracker is capable of operating closer to $\mathrm{S/N}_\phi = 1$, the phase tracking S/N limit \citep{Colavita+1999, Colavita+2010, Woillez+2012, Lacour+2019}.

Even though the updated GRAVITY on-axis fringe tracker limiting magnitude, at $\mathrm{K} = \SI{8}{\mag}$ on the ATs, is only one magnitude shy of the UT limit, the performance on the UT is still ten times better\footnote{The UT transmission is twice as low as the AT's, but the collecting power is 20 times as high.} than on the ATs in terms of collected photons and the associated effect on interferometric observables.
This observation only illustrates an issue with the UTs, probably resulting from a lower median Strehl ratio and stronger telescope vibrations.

\subsection{Resilience to degraded seeing conditions}\label{SSec:Resilience}

\begin{figure}
    \centering
    \includegraphics[width=0.8\linewidth]{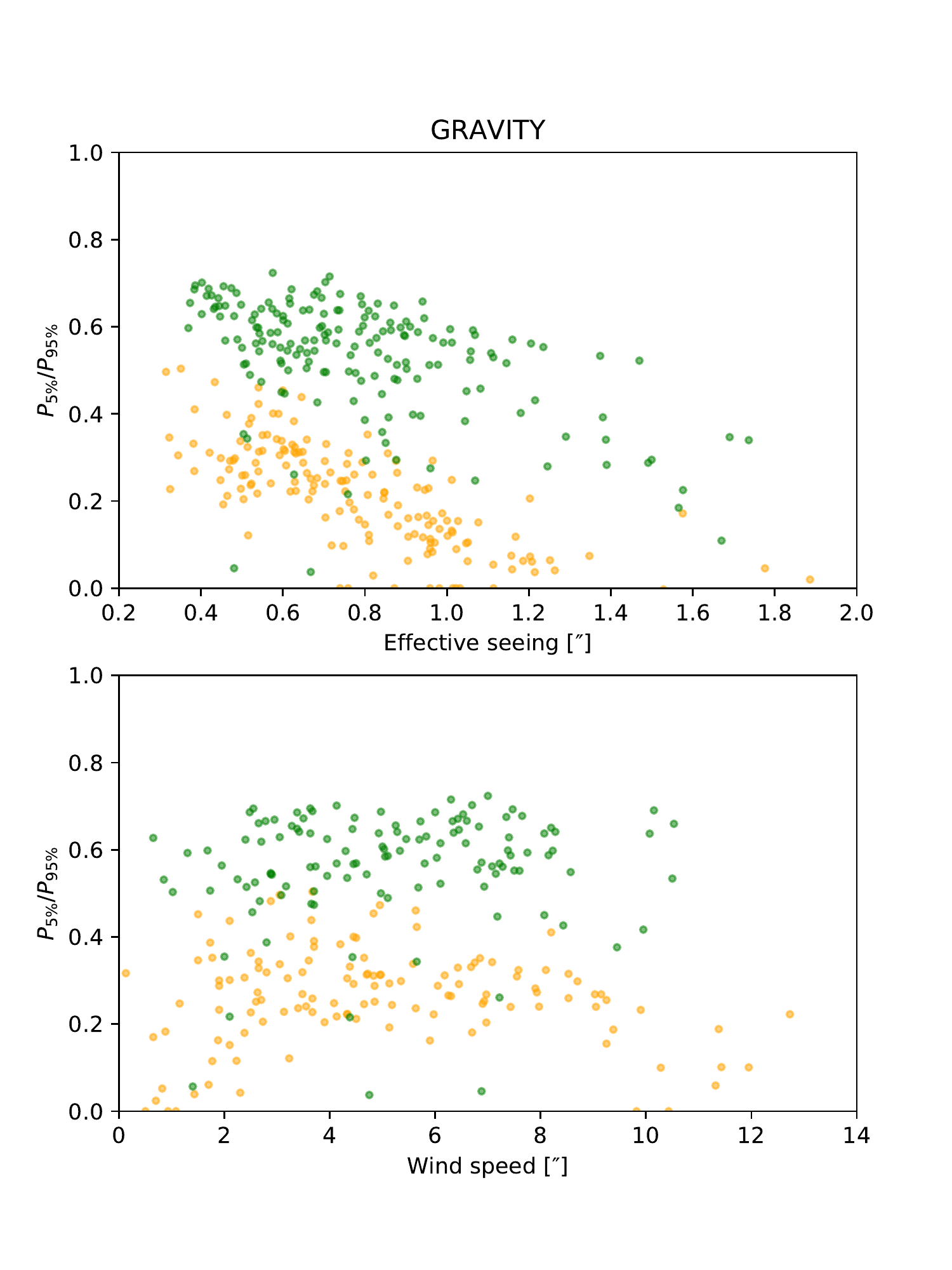}
    \caption{Comparison of GRAVITY fringe tracker injection stability between STRAP and NAOMI. \textbf{Top}: Injection stability vs. effective seeing for wind speeds in the \SIrange{3}{9}{\meter\per\second} range. \textbf{Bottom}: Injection stability vs. wind speed for a better-than-median effective seeing. NAOMI obtains in \SI{1.5}{\arcsec} seeing the same performance as STRAP in \SI{0.5}{\arcsec} seeing. NAOMI does not exhibit performance loss when the wind speed is below \SI{2}{\meter\per\second}.}
    \label{Fig:SeeingWindSpeed}
\end{figure}

One last objective of the NAOMI project was to make the AT array more resilient to degraded seeing conditions.
This includes larger-than-median free atmosphere and ground-layer seeing degradations, as well as low wind ($<\SI{2}{\meter\per\second}$) nights with severe dome seeing.
We therefore explored correlations between the injection stability metric $P_{5\%}/P_{95\%}$ and seeing or wind speed.
As illustrated in Fig.~\ref{Fig:SeeingWindSpeed} for the GRAVITY fringe tracker, NAOMI fulfils its objectives.
At wind speeds below \SI{2}{\meter\per\second}, the severely degraded injection with STRAP ($P_{5\%}/P_{95\%} < 0.2$) is replaced by a stable wind-speed-independent injection ($P_{5\%}/P_{95\%} \sim 0.5$).
In $\sim\SI{1.25}{\arcsecond}$ seeing conditions, NAOMI delivers the same performance level as was previously achieved at $\sim\SI{0.5}{\arcsecond}$ seeing by STRAP.

\section{Conclusions}

The deployment of the NAOMI adaptive optics system on the VLTI is a perfect illustration of the benefit of high and stable Strehl ratio for ground-based optical interferometry in general and fringe tracking in particular.
Beyond the transmission improvement, the VLTI can now operate with the ATs in degraded seeing conditions, which increases the amount of scientifically exploitable time on the interferometer: a previously unusable AT-level seeing threshold of \SI{1.4}{\arcsec} would correspond to a $+15\%$ increase in usable time, based on Paranal seeing statistics.
The fringe tracking performance of GRAVITY has improved as well: a high Strehl ratio allows continuous measurements of the phase without fringe jumps.
This will become even more evident when the K-band GRAVITY fringe tracker is used to co-phase the L/M/N-band MATISSE instrument.
The improved Strehl ratio also supports extending the VLTI toward shorter wavelengths.
In the J band and in median seeing conditions, the Strehl ratio should have increased from $\sim8\%$ to $\sim40\%$, which is the level previously obtained with STRAP in the K band.
Beyond the new scientific window, a J-band extension also represents an increase in angular resolution by $+25\%$.
In addition, the availability of AO-corrected telescopes allows non-spatially filtered instrument concepts to be studied, which might enable significantly higher throughputs.

However, this improved performance is only available for objects that meet the bright regime requirement of $R < \SI{12}{\mag}$.
For the current level of performance of the near-infrared VLTI instruments on the \SI{1.8}{\meter} ATs, this is not a problem most of the time.
The colours that bring both the AO and the instruments to their respective limits are already red at $R-H|K = +\SI{3}{\mag}$.
The adaptive optics is therefore never the limiting factor for surface temperatures down to $\sim\SI{3750}{\kelvin}$ ($\mathrm{M}0$ spectral type).
The situation is very different on the \SI{8}{\meter} UTs.
An AO system operating at the same sub-aperture size as NAOMI, that is, at higher order than the current MACAO, would have a colour limit at $R-H|K = +\SI{0}{\mag}$, which is equivalent to a surface temperature of $\sim\SI{9790}{\kelvin}$ ($\mathrm{A}0$ spectral type).
This is a quantitative way of expressing the feeling that the AO system is always the first thing to fail on the UTs.
This shows that reproducing the same kind of improvement on the \SI{8}{\meter} telescopes will certainly require laser guide-star adaptive optics \citep{Beckers1990}.

\begin{acknowledgements}
    This research made use of Astropy,\footnote{http://www.astropy.org} a community-developed core Python package for Astronomy \citep{astropy+2013, astropy+2018}. S.Z-F. acknowledges support from Iniciativa Cient\'ifica Milenio via N\'ucleo Milenio de Formaci\'on Planetaria. S.Z-F acknowledges financial support from {CONICYT} via PFCHA/Doctorado Nacional/2018-21181044 and the European Southern Observatory via its studentship program.
\end{acknowledgements}


\bibliographystyle{aa}
\bibliography{naomi}

\end{document}